\documentclass[secthm,seceqn,amsthm,ussrhead,eqno]{amsart}
\usepackage{amsmath,latexsym}
\usepackage[psamsfonts]{amssymb}
\usepackage{times}
\usepackage[mathcal]{euscript}
\numberwithin{equation}{section}

\newcommand{\bbT}{\mathbb T}

\renewcommand{\epsilon}{\varepsilon}

\newcommand{\be}{\begin{equation}}
\newcommand{\ee}{\end{equation}}
\newcommand{\no}{\nonumber}

\newcommand{\C}{\mathbb{C}}

\newcommand{\R}{\mathbb{R}}
\renewcommand{\S}{\mathbb{S}}
\newcommand{\T}{\mathbb{T}}

\newcommand{\Z}{\mathbb{Z}}

\newcommand{\cU}{{\mathcal U}}

\renewcommand{\Im}{{\ensuremath{\mathrm{Im}}}}
\renewcommand{\Re}{{\ensuremath{\mathrm{Re}}}}

\renewcommand{\det}{\mathop{\mathrm{det}}}

{\bf}{\it}
\newtheorem{theorem}{Theorem}[section]
\newtheorem{lemma}[theorem]{Lemma}
\newtheorem{corollary}[theorem]{Corollary}
\newtheorem{hypothesis}[theorem]{Hypothesis}
\newtheorem{definition}[theorem]{Definition}
\newtheorem{proposition}[theorem]{Proposition}
\newtheorem{remark}[theorem]{Remark}

\date{\today}

\begin{document}
\title[On the essential and discrete spectrum of a model operator...]
{On the essential and discrete spectrum of a model operator
related to three-particle discrete Schr\"odinger operators}

\author{Sergio  Albeverio$^{1,2,3}$, Saidakhmat  N. Lakaev$^{4,5}$,
  Ramiza Kh. Djumanova $^{5}$}

\address{$^1$ Institut f\"{u}r Angewandte Mathematik,
Universit\"{a}t Bonn, Wegelerstr. 6, D-53115 Bonn\ (Germany)}

\address{
$^2$ \ SFB 611, \ Bonn, \ BiBoS, Bielefeld - Bonn;}
\address{
$^3$ \ CERFIM, Locarno and Acc.ARch,USI (Switzerland) E-mail
albeverio@uni.bonn.de}

\address{
{$^4$ Samarkand State University, University Boulevard 15, 703004,
Samarkand (Uzbekistan)} \ {E-mail: slakaev@mail.ru }}

\address{$^5$ {Samarkand division of Academy of sciences of
Uzbekistan (Uzbekistan)}} \maketitle

\begin{abstract}
A model operator $H$ corresponding to a three-particle discrete
Schr\"odinger operator on a lattice $\Z^3$ is studied. The
essential spectrum is described via the spectrum of  two
Friedrichs models with parameters $h_\alpha(p),$ $\alpha=1,2,$ $p
\in \T^3=(-\pi,\pi]^3.$ The following results are proven:

1) The operator $H$ has a finite number of eigenvalues lying below
the bottom of the essential spectrum in any of the following
cases: (i) both operators $h_\alpha(0),\,\alpha=1,2,$  have a zero
eigenvalue; (ii) either $h_1(0)$ or $h_2(0)$ has a zero
eigenvalue.

2) The operator $H$ has infinitely many eigenvalues lying below
the bottom
  and accumulating at the bottom of the essential spectrum, if both operators
$h_\alpha(0),\alpha=1,2,$ have a zero energy resonance.
\end{abstract}



\section{Introduction}

One of the remarkable results in the spectral analysis for
continuous and discrete three-particle Schr\"odinger  operators is
the Efimov effect: if in a system of three-particles, interacting by
means of short-range pair potentials, none of the three two-particle
subsystems has bound states with negative energy, but at least two
of them have a resonance with zero energy, then this three-particle
system has an infinite number of three-particle bound states with
negative energy, accumulating at zero.

This effect was first discovered by Efimov \cite{Ef}. Since then
this problem has been studied in many  works
\cite{AHW,ALkM,AN,DelFT,FaMe,OvSi,Sob,Tjfa91,Tmsj94,Yafms74}. A
rigorous mathematical proof of the existence of Efimov's  effect
was originally carried out by Yafaev in \cite{Yafms74} and then in
\cite{OvSi,Tjfa91,Sob,Tmsj94}.

In models of solid physics \cite{GS,FIC,Mat,Mog,ReSiIII,Yaflnm00}
and also in lattice quantum field theory \cite{MaMi} discrete
Schr\"{o}dinger operators are considered, which are lattice
analogues of the three-particle Schr\"{o}dinger operator in a
continuous space. The presence of Efimov's effect for these
operators was proved in \cite{ALzM,Ltmf91,Lfa93,LAfa99}.

In \cite{ALzM} a system of three arbitrary quantum particles on
the three-dimensional lattice $\Z^3$ interacting via zero-range
pair attractive potentials  has been considered.

Let us denote by $\tau_{ess}(K)$ the bottom of essential spectrum of
the three-particle discrete Schr\"{o}dinger operator $H(K),\, K \in
{\bbT}^3,$  and by $N(K,z)$ the number of eigenvalues below
 $z \leq \tau_{ess}(K).$

Let us shortly recall the main results of \cite{ALzM}:

(i) For the number $N(0,z)$ the limit result
$$
\lim_{z \to -0} \frac{N(0,z)}{|\log|z||}={\cU}_0,\, (0<{\cU}_0<
\infty)
$$
holds.

 (ii) For any  $K \in U^{0}_{\delta}(0)=\{p\in \T^3:0<|p|< \delta\}$ the
 number $N(K,\tau_{ess}(K))$ is finite and the following
limit result
$$
\lim_{|K| \to 0} \frac{N(K,0)}{|\log|K||}=2{\cU}_0
$$
holds (see \cite{ALzM} for details).

In the present paper a model operator $H$ corresponding  to the
three-particle Schr{\"o}dinger operator on the  lattice $\Z^3$
 acting in the Hilbert
space $L_2(({\T}^3)^2)$ is considered. Here the role of the
two-particle discrete Schr{\"o}dinger operators is  played by a
family of Friedrichs  models with parameters
$h_\alpha(p),\,\alpha=1,2,\,p \in \T^3.$

We precisely describe the location and structure of the essential
spectrum of $H$ via the spectrum of $h_\alpha(p),\,\alpha=1,2,\,p
\in \T^3.$

Furthermore under some natural conditions on the family of the
operators $h_\alpha(p),\,\alpha=1,2,\,p \in \T^3,$
 we obtain the following results:

$(a)$ The operator $H$ has a finite number of eigenvalues lying
below the bottom of the essential spectrum in the following two
cases: (i) both operators $h_\alpha(0),\,\alpha=1,2,$ have a zero
eigenvalue; (ii) either $h_1(0)$ or $h_2(0)$ has a zero
eigenvalue.

$(b)$ The operator $H$ has infinitely many eigenvalues lying below
the bottom
  and accumulating at the bottom of the essential spectrum, if the operators
$h_\alpha(0),\alpha=1,2,$ have a zero energy resonance.

Moreover for the number $N(z)$ of eigenvalues of $H$ lying below
$z<0$ the following limit exists
\begin{equation*} \label{asym.K}
\lim\limits_{z \to -0}\frac{N(z)}{|\log |z||}={\cU}_0 \,(0<{\cU}_0
<\infty).
\end{equation*}

We remark that the assertion (b) is similar to the case of the
three-particle continuous and discrete Schr\"odinger operators and
the assertion (a) is surprising and similar assertions has not
been proved for the three-particle Schr\"odinger operators on
$\R^3$ and $\Z^3.$

The plan of  this paper is as follows:

 Section 1 is an
introduction to the whole work. In section 2 the model operator
$H$ is introduced in the Hilbert space $L_2((\T^3)^2)$ and the
main results of the present paper are formulated. In Section 3 we
study some spectral properties of $h_\alpha(p),\,\alpha=1,2,\,p
\in \T^3$.
 In section 4 we obtain an analogue of the Faddeev-Newton type
integral equation for the eigenfunctions of the model operator and
precisely describe the location and the structure of the essential
spectrum of $H$ (Theorem \ref{ess}). In this section we prove an
analogue of the Birman-Schwinger principle for $H$  and the part
$(i)$ of Theorem \ref{fin}. In section 6 we prove the part $(ii)$
of Theorem \ref{fin}. Some technical material is collected in
Appendices A, B.

Throughout the present paper we adopt the following conventions:
For each $\delta>0$ the notation $U_{\delta}(0) =\{p\in
{\bbT}^3:|p|<\delta \}$ stands for a $\delta$-neighborhood of the
origin.

The subscript $\alpha$ (and also $\beta$) always is equal to $1$
or $2$ and $\alpha\neq\beta$ and $\T^3$ denotes  the
three-dimensional torus, the cube $(-\pi,\pi]^3$ with
appropriately  identified sides. Throughout the paper the torus
$\T^3$
 will always be considered as an
abelian group with respect to the addition and multiplication by
real numbers
 regarded as operations
on $\R^3$ modulo $(2\pi \Z)^3$.

Denote by $L_2(\Omega)$ the Hilbert space of square integrable
functions defined on a measurable set $\Omega \subset \R^n,$ and by
$L_2^{(2)}(\Omega)$ the Hilbert space of two-component vector
functions $f=(f_1,f_2)$,\,$f_\alpha \in L_2(\Omega),\,\alpha=1,2.$

\section{Three particle model operator and
statement of  results}


Let us consider the operator $H$ acting in the Hilbert space
$L_2((\T^3)^2)$ by
\begin{equation}\label{oper H}
H=H_0-\mu_1V_1-\mu_2V_2,
\end{equation}
where
\begin{align*}
&(H_0f)(p,q)=u(p,q)f(p,q), \quad f\in L_2((\T^3)^2),\\
&(V_1f)(p,q)=\varphi_1(p)\int_{{\T}^3}\varphi_1(t) f(t,q)dt, \quad
f\in L_2((\T^3)^2),\\
&(V_2f)(p,q)=\varphi_2(q)\int_{{\T}^3}\varphi_2(t) f(p,t)dt, \quad
f\in L_2((\T^3)^2).
\end{align*}

  Here $u(p,q)$ and $\varphi_\alpha(p),\,\, \alpha=1,2,$ are
real-analytic functions defined on $({\T}^3)^2$ and ${\T}^3,$
respectively, and  $\mu_{\alpha},\alpha=1,2,$ are positive  real
numbers.

Under these assumptions the operator $H$ defined by \eqref{oper H}
is bounded and self-adjoint.

Throughout this paper we assume  the following additional
hypothesis.

\begin{hypothesis}\label{hypoth u} The real-analytic function $u(p,q)$ on
$({\T }^3)^2$ is even with respect to $(p,q),$ has a unique
non-degenerate zero minimum at the point $(0,0)\in ({\T}^3)^2$ and
there exists a positive definite matrix $U$ and real numbers
$l,l_1, l_2 \,(l_1,l_2>0,l\not=0)$   such that $$ \left(
\frac{\partial^2 u(0,0)}{\partial p^{(i)} \partial p^{(j)}}
\right)_{i,j=1}^3= l_1 U,\,\, \left( \frac{\partial^2
u(0,0)}{\partial p^{(i)} \partial q^{(j)}} \right)_{i,j=1}^3= l
U,\,\, \left( \frac{\partial^2 u(0,0)}{\partial q^{(i)} \partial
q^{(j)}} \right)_{i,j=1}^3= l_2 U.
$$
\end{hypothesis}
\begin{hypothesis}\label{hyp.varphi} The real-analytic function
 $\varphi_{\alpha}(p),\alpha=1,2,$  is either even or odd on
$\T^3$.
\end{hypothesis}

Set
 $$ u_p^{(1)}(q)=u(q,p),\,\,u_p^{(2)}(q)=u(p,q).$$

By Hypotheses \ref{hypoth u} and \ref{hyp.varphi} the integral
\begin{equation}\label{Lamb}
\int\limits_{{\T}^3}
\frac{\varphi_\alpha^2(t)dt}{u_p^{(\alpha)}(t)}
\end{equation}
is finite and hence we can define continuous function on $\T^3,$
which will be denotes  $\Lambda_{\alpha}(p).$

\begin{remark} Since the
function $u(p,q)$ has  a unique non degenerate minimum at the
point $(0,0)\in ({\T}^3)^2$ the function $\Lambda_{\alpha}(p)$ is
positive. In particular, if $\varphi_\alpha(0)=0$ then
$\Lambda_{\alpha}(p)$ is a twice continuously differentiable
function at the point $p=0$ (see proof of Lemma \ref{examp}).
\end{remark}
\begin{hypothesis}\label{hypothD}

 (i) For any $p\in
\T^3,\,p\neq 0$ the function $\Lambda_{\alpha}(\cdot)$ satisfies
$\Lambda_{\alpha}(p)<\Lambda_{\alpha}(0).$\\ (ii) If
$\varphi_\alpha(0)=0,$ then $\Lambda_{\alpha}(p)$ has a
non-degenerate maximum at $p=0.$
\end{hypothesis}
\begin{remark}
Let Hypotheses \ref{hypoth u} and \ref{hyp.varphi} be fulfilled
and $\varphi_\alpha(0)\neq 0$. Then it is easy to show that the
assertion $(i)$ of Hypothesis \ref{hypothD} is fulfilled, that is,
the inequality $\Lambda_{\alpha}(p)<\Lambda_{\alpha}(0)$ holds for
all sufficiently small nonzero $p\in \T^3$ (see Corollary
\ref{razl.lemma.natijasi.0}).
\end{remark}
\begin{remark}\label{examp0} There are a function $u(p,q)$
and either even or odd functions $\varphi_\alpha(p),\alpha=1,2,$ so
that Hypothesis \ref{hypothD} is  fulfilled (see Appendix B).
\end{remark}

Set
\begin{equation*}\label{mu.alpha}
  \mu_\alpha^0=\Big( \int_{{\T}^3}
\varphi_\alpha^2(t) (u_0^{(\alpha)}(t))^{-1}dt \Big)^{-1}\quad
\mbox{and}\quad  M=\max_{p,q\in \T^3}u(p,q).
\end{equation*}

To study  spectral properties of the operator $H$ we introduce the
following two families of bounded self-adjoint operators
(Friedrichs model) $\{h_{\alpha}(p),\, \alpha=1,2,\, p\in
{\T}^3\},$ acting in $L_2(\T^3)$ by
\begin{equation}\label{h_alpha}
 h_\alpha(p)=h_\alpha^{0}(p)-\mu_\alpha v_\alpha,
 \end{equation}
 where
\begin{align*}
 (h^0_\alpha(p)f)(q)=u_p^{(\alpha)}(q)f(q),\quad
 f\in L_2(\T^3),\,\\
 (v_{\alpha}f)(q)=\varphi_\alpha(q)\int\limits_{\T^3}
\varphi_\alpha(t)f(t)dt,\quad
 f\in L_2(\T^3).
 \end{align*}

For the definition  of Friedrichs model and  the study spectrum
and resonances in this model see
\cite{Frie,Fadtms64,Lfa83,Yafmst92}.

Let $\sigma_d(h_\alpha(p))$ be the discrete spectrum of
$h_\alpha(p),\,p\in {\T^3},$ and
\begin{align*}
a_\alpha\equiv\inf\cup_{p\in {\T}^3} \sigma_d(h_\alpha(p)),\,\,
b_\alpha\equiv\sup\cup_{p\in {\T}^3}
\sigma_d(h_\alpha(p)),\,\alpha=1,2.
\end{align*}

The main results of the present paper are as follows:
\begin{theorem}\label{ess}
Assume Hypothesis \ref{hypoth u} and \ref{hypothD} are fulfilled.\\
 (i) Let $\mu_\alpha >\mu_\alpha^{max},\,\alpha=1,2,$ then
\begin{equation*}
{\sigma}_{ess}(H)=[a_1,b_1]\cup [a_2,b_2]\cup [0, M]\quad
\mbox{and}\quad b_\alpha<0,\,\alpha=1,2.
\end{equation*}
(ii) Let $ \mu_\alpha^{max}\geq  \mu_\alpha>
\mu_\alpha^0,\,\alpha=1,2$, then
\begin{equation*}
{\sigma}_{ess}(H)= [a, M]\quad \mbox{and}\quad a=min\{a_1,a_2\}<0.
\end{equation*}
(iii) Let $   \mu_\alpha^0\geq \mu_\alpha >0,\,\alpha=1,2$, then
\begin{equation*}
{\sigma}_{ess}(H)= [0,M].
\end{equation*}
\end{theorem}


Let $C(\T^3)$ be the Banach space of continuous functions on $\T^3.$

\begin{definition}\label{resonance0} Let Hypotheses \ref{hypoth u}
be fulfilled.
 The operator $h_\alpha (0),\,\alpha=1,2,$ is said to
have a zero energy resonance if the number  $1$ is an eigenvalue
of the integral operator given by $$
(\mathrm{G}_\alpha\psi_\alpha)(q)=\mu_\alpha \varphi_\alpha(q),
\int\limits_{{\T}^3}
\frac{\varphi_\alpha(t)\psi_\alpha(t)dt}{u_0^{(\alpha)}(t)},\,\,\psi_\alpha\in
{C(\T^3)}$$ and  $\varphi_\alpha(0)\neq 0.$
\end{definition}

\begin{theorem}\label{fin}  Assume Hypotheses \ref{hypoth u},\ref{hyp.varphi}
and \ref{hypothD} are fulfilled and $\mu_\alpha =\mu_\alpha^0,\,\alpha=1,2$.\\
 (i)  Let either
$\varphi_1(0)=\varphi_2(0)=0$ or
$\varphi_1(0)=0,\varphi_2(0)\neq0$
 or $\varphi_1(0)\neq 0,\varphi_2(0)=0$.  Then the operator $H$ has a finite
number of eigenvalues outside of the essential spectrum.\\
 (ii) Let $\varphi_\alpha(0)\neq 0$ for all
$\alpha=1,2.$  Then the discrete spectrum of $H$ is infinite and
the function $N(z)$  obeys the relation
\begin{equation} \label{asym.K}
\lim\limits_{z \to -0}\frac{N(z)}{|\log |z||}={\cU}_0 \,(0<{\cU}_0
<\infty).
\end{equation}
\end{theorem}
\begin{remark} The constant ${\cU}_0$ does not depend on the
functions $\varphi_\alpha(p),\,\alpha=1,2,$  and is given as a
positive function depending only on the ratios $\frac
{l_\alpha}{l},\alpha=1,2$.
\end{remark}

\begin{remark} The conditions $\mu_\alpha =\mu_\alpha^0$ and
$\varphi_\alpha(0)\neq 0,\,\alpha=1,2,$ (resp.
$\varphi_\alpha(0)=0)$ means that the operator $h_\alpha(p)$ has a
zero-energy resonance (resp. zero eigenvalue) (see Lemma
\ref{resonance} (resp. Lemma \ref{zeroeigen})).
\end{remark}
\begin{remark} Clearly, the infinite cardinality of the negative discrete spectrum of $H$
follows automatically from the positivity of ${\cU}_{0}.$
\end{remark}
\begin{remark}\label{techn}
We note that the assumptions for the functions $u$ and
$\varphi_i,\,i=1,2,$ are far from the precise, but we will not
develop this point here.
\end{remark}

\section{Spectral properties of the operator $h_\alpha(p)$}
In this section we study  some spectral properties of the operator
$h_\alpha(p),\,\,p \in \T^3$  given by \eqref{h_alpha}.

The perturbation $v_\alpha$ of the multiplication operator
$h_\alpha^0(p)$ is a one-dimensional self-adjoint integral operator.
Therefore in accordance with invariance of the absolutely continuous
spectrum under  trace class perturbations the absolutely continuous
spectrum  of the operator $h_\alpha(p)$ fills  the following
interval on the real axis:
$$
\sigma_{ac}(h_\alpha(p))=[m_\alpha(p),M_\alpha(p)],$$ where the
numbers $m_\alpha(p)$ and $M_\alpha(p)$ are defined by
\begin{equation*}
m_\alpha(p)=\min_{q\in {\T}^3}u^{(\alpha)}_p(q)
,\,\,M_\alpha(p)=\max_{q\in {\T}^3} u^{(\alpha)}_p(q).
\end{equation*}

Let ${\C}$ be the field of complex numbers. For any $p \in \T^3$ and
$z{\in } {\C} { \setminus } \sigma_{ac}(h_\alpha(p))$ we define the
function (the Fredholm determinant
 associated with the operator $h_\alpha(p)$)
\begin{equation*}\label{det}
\Delta_{\mu_\alpha}(p,z)=1-\mu_\alpha \int\limits_{{\T}^3}
\frac{\varphi_\alpha^2(t)dt}{u_p^{(\alpha)}(t)-z}.
\end{equation*}

Note that  ${\Delta}_{\mu_\alpha}( p, z)$ is real-analytic in
$\T^3\times ({\C} { \setminus } \sigma_{ac}(h_\alpha(p))).$

Denote by $r_\alpha^{0}(p,z)=(h^0_\alpha( p)-z)^{-1}$ the
resolvent  of the operator $h^0_\alpha(p)$, that is, the
multiplication operator by the function
$(u_p^{(\alpha)}(t)-z)^{-1}.$

\begin{lemma}\label{delta=0}
For all $\mu_\alpha>0$ and $p\in \T^3$ the following statements are
equivalent:\\ (i) The operator $h_{\alpha}(p)$ has an eigenvalue $z
\in {\C} \setminus \sigma_{ac}(h_\alpha(p))$ below the bottom of the
continuous spectrum.\\ (ii) $\Delta_{\mu_\alpha}(p,z)=0$, $z \in
{\C} \setminus \sigma_{ac}(h_\alpha(p)).$\\
 (iii) $\Delta_{\mu_\alpha}(p,z')<0$ for some
$z'\leq m_\alpha(p).$
\end{lemma}
\begin{proof}
The number $z \in {\C} \setminus \sigma_{ac}(h_\alpha(p))$ is an
eigenvalue of $h_\alpha(p)$ if and only if
 (by the Birman-Schwinger principle)
$\lambda=1$ is an eigenvalue of the operator
\begin{equation*}
G_{\mu_\alpha}(p,z)=\mu_\alpha v^{\frac{1}{2}}_\alpha
r^0_\alpha(p,z)v^{\frac{1}{2}}_\alpha.
\end{equation*}

Since the operator $v^{\frac{1}{2}}_\alpha $ is of the form $$
(v^{\frac{1}{2}}_\alpha
f)(q)=||\varphi_\alpha||^{-1}\varphi_\alpha(q)\int_{\T^3}
\varphi_\alpha(t)f(t)dt,\,f \in L_2(\T^3) $$ the operator
$G_{\mu_\alpha}(p,z)$ has the form
\begin{equation*}
(G_{\mu_\alpha}(p,z)f)(q)=\frac{\mu_\alpha {\Lambda}_\alpha(p,z)
}{||\varphi_\alpha||^2} \varphi_\alpha(q)\int_{\T^3}
\varphi_\alpha(t)f(t)dt,\,f \in L_2(\T^3),
\end{equation*}
where
\begin{equation*}
\Lambda_\alpha(p,z)=\int_{\T^3}
\frac{\varphi_\alpha^2(t)}{u_p^{(\alpha)}(t)-z}dt.
\end{equation*}

According to Fredholm's theorem  the number $\lambda=1$ is an
eigenvalue for the operator $G_{\mu_\alpha}(p,z)$ if and only if
$$
  1-\mu_\alpha \Lambda_\alpha(p,z)=0,
\quad \mbox{that is,}\quad  \Delta_{\mu_\alpha}(p,z)=0. $$ The
equivalence of $(i)$ and $(ii)$ is proven.

Now we prove the equivalence of $(ii)$ and $(iii)$.
 Let $\Delta_{\mu_\alpha}(p,z_0)=0$ for some $z_0 \in
\C\setminus  \sigma_{ac}(h_\alpha(p))$. The operator $h_\alpha(p)$
is self-adjoint and $(i)$ and $(ii)$ is equivalent hence  $z_0$ is a
real number. Since $\Delta_{\mu_\alpha}(p,z)>1$ for all
$z>M_\alpha(p)$, we have $z_0\in (-\infty,m_\alpha(p)).$
 Since for any $p\in
{\T}^3$ the function $\Delta_{\mu_\alpha}(p,z)$ is decreasing in
$z\in (-\infty,m_\alpha(p))$ we have $\Delta_{\mu_\alpha}(p,z')<
\Delta_{\mu_\alpha}(p,z_0)=0$ for some $ z_0<z'<m_\alpha(p).$

Now we suppose that $\Delta_{\mu_\alpha}(p,z')<0$ for some $z'
\leq m_\alpha(p).$ Since for any $p\in {\T}^3$ $\lim\limits_{z\to
-\infty} \Delta_{\mu_\alpha}(p,z)=1$  and
$\Delta_{\mu_\alpha}(p,z)$ is continuous in $z\in
(-\infty,m_\alpha(p))$, we obtain that there exists $z_0\in
(-\infty;z')$ such that $\Delta_{\mu_\alpha}(p,z_0)=0.$ This
completes the proof.
\end{proof}

Since the function $\Delta_{\mu_\alpha}(0,\cdot)$ is decreasing on
$(-\infty,0)$ and the function $u^{(\alpha)}_0(q)$ has a unique
non-degenerate minimum at $q=0$ (see Lemma \ref{minimum}) by
dominated convergence there exist the finite limit
$$
\Delta_{\mu_\alpha}(0,0)=\lim_{z\to -0} \Delta_{\mu_\alpha}(0,z).
$$

\begin{lemma}\label{resonance}   Let Hypotheses \ref{hypoth u} be fulfilled.
The following statements are equivalent:\\
(i) the operator $h_{\alpha}(0)$ has a zero energy resonance.\\
(ii)  $\varphi_\alpha(0)\neq 0$  and
$\Delta_{\mu_\alpha}(0,0)=0.$\\
(iii) $\varphi_\alpha(0)\neq 0$  and $\mu_\alpha= \mu_\alpha^{0}.$
\end{lemma}
  \begin{proof} Let the operator $h_\alpha(0)$ have
a zero energy  resonance for some $\mu_\alpha>0$. Then by
 Definition \ref{resonance0} the equation
\begin{equation*}
\psi_\alpha(q)=\mu_\alpha \varphi_\alpha(q) \int\limits_{{\T}^3}
\frac{\varphi_\alpha(t)\psi_\alpha(t)dt}{u_0^{(\alpha)}(t)},\,\,\psi_\alpha\in
{C(\T^3)}
 \end{equation*}
 has a simple solution $\psi_\alpha(q)$ in $C({\T^{3}})$.

One can  check that this solution is equal to the  function
$\varphi_\alpha(q)$ (up to a constant factor).
 Therefore we see that
 $$
\varphi_\alpha(q)=\mu_\alpha
 \varphi_\alpha(q) \int\limits_{{\T}^3}
\frac{\varphi^2_\alpha(t)dt}{u_0^{(\alpha)}(t)}
 $$
and hence
$$ \Delta_{\mu_\alpha}(0,0)= 1-\mu_\alpha
\int\limits_{{\T}^3}
\frac{\varphi^2_\alpha(t)dt}{u_0^{(\alpha)}(t)} =0
$$ and so
$\mu_\alpha=\mu_\alpha^{0}$.

 Let for some $\mu_\alpha>0$ the equality
 $$\Delta_{\mu_\alpha}(0,0)=
1-\mu_\alpha \int\limits_{{\T}^3}
\frac{\varphi^2_\alpha(t)dt}{u_0^{(\alpha)}(t)}
 $$
 hold and consequently $\mu_\alpha=\mu_\alpha^{0}$.
Then  only the function $\varphi_{\alpha}(q) \in  C({\T}^{3})$ is
a solution of the equation
 $$
\psi_{\alpha}(q)=\mu_\alpha \varphi_\alpha(q)\int\limits_{{\T}^3}
\frac{\varphi_\alpha(t)\psi_{\alpha}(t)dt}{u_0^{(\alpha)}(t)}
 ,$$
  that is, the
operator $h_\alpha(0)$ has a zero energy resonance.
\end{proof}

\begin{lemma}\label{zeroeigen} Let Hypotheses \ref{hypoth u}
 be fulfilled.
  The following statements are equivalent:\\
(i) the operator $h_{\alpha}(0)$ has a zero eigenvalue.\\
(ii)  $\varphi_\alpha(0)=0$  and
$\Delta_{\mu_\alpha}(0,0)=0.$\\
(iii) $\varphi_\alpha(0)=0$ and $\mu_\alpha= \mu_\alpha^{0}.$
\end{lemma}
\begin{proof}
Suppose $f\in L_2(\T^3)$ is an eigenfunction of the operator
$h_\alpha(0)$ associated with the zero eigenvalue. Then $f$
satisfies the equation
\begin{equation}\label{h=0}
u_0^{(\alpha)}(q)f(q)-\mu_\alpha \varphi_\alpha(q)
\int\limits_{{\T}^3} \varphi_\alpha(t)f(t)dt=0.
\end{equation}
From \eqref{h=0} we find that $f,$ except for an arbitrary factor,
is given by
\begin{equation}\label{eigenunction}
f(q)=\frac{\varphi_\alpha(q)}{u_0^{(\alpha)}(q)},
\end{equation}
and from \eqref{h=0} we derive the equality
$\Delta_{\mu_\alpha}(0,0)=0$ and so $\mu_\alpha=\mu^0_\alpha.$

Since the function $u_0^{(\alpha)}(q)$ has a non-degenerate minimum
at the point $q=0$ (see Lemma \ref{minimum}) from
\eqref{eigenunction} we have $\varphi_\alpha(0)=0.$

Substituting the expression \eqref{eigenunction} for the $f$ to the
\eqref{h=0} we get the equality
\begin{equation*}
\varphi_\alpha(q)=\mu_\alpha \varphi_\alpha(q)
\int\limits_{{\T}^3}
\frac{\varphi_\alpha^2(t)dt}{u_0^{(\alpha)}(t)}.
 \end{equation*}

Hence $\Delta_{\mu_\alpha}(0,0)=0$ and so
$\mu_\alpha=\mu^0_\alpha.$

Let $\varphi_{\alpha}(0)=0$ and  $\Delta_{\mu_\alpha}(0,0)=0$ then
 $\mu_\alpha=\mu^0_\alpha$ and the function
$$
 f(q)=\frac{\varphi_{\alpha}(q)}{u^{(\alpha)}_0(q)}
$$
  obeys the
equation
$$ h_\alpha(0)f=0$$
and
$$
f\in L_2(\T^3).
$$

Indeed, the functions $u^{(\alpha)}_0(q)$ and
$\varphi_{\alpha}(q)$ are analytic on $\T^3$ and the function
$u^{(\alpha)}_0(q)$  has a unique non-degenerate minimum at the
origin, hence  $$ f(q)=\frac{\varphi(q)}{u^{(\alpha)}_0(q)} \in
L_2(\T^3)$$ if and only if $\varphi_{\alpha}(0)=0$.
 \end{proof}

Set
\begin{equation*}\label{mu.max}
\mu_\alpha^{max}= \max_{p\in\T^3}\Lambda_\alpha^{-1}(p,0).
\end{equation*}

\begin{lemma}\label{aaa} Let Hypotheses \ref{hypoth u} and
\ref{hypothD} be fulfilled.\\
(i) Let $\mu_\alpha>\mu_\alpha^{max}.$ Then for any $p \in \T^3$
the operator $h_\alpha(p)$ has a unique negative eigenvalue.\\
(ii) Let $\mu_\alpha^{max}\geq \mu_\alpha>\mu_\alpha^0.$ Then there
exists a non void open set $D_{\mu_\alpha}\subset \T^3$ such that
for any $p \in D_{\mu_\alpha}$ the operator $h_\alpha(p)$ has a
unique negative eigenvalue and for $p \in \T^3\setminus
D_{\mu_\alpha}$ the operator $h_\alpha(p)$ has no negative
eigenvalues.\\
(iii) Let $0<\mu_\alpha\leq \mu_\alpha^0.$ Then for any $p \in
\T^3$ the operator $h_\alpha(p)$ has no negative eigenvalues.
\end{lemma}
\begin{proof}$(i)$ Let $\mu_\alpha>\mu_\alpha^{max}.$ Since $\T^3$ is a compact
set and the function $\Lambda^{-1}_\alpha(p,0)$ is the continuous on
$\T^3$ by the definition of $\mu_\alpha^{max}$ for all $p \in \T^3$
we get the inequalities
\begin{equation*}
\Lambda_\alpha^{-1}(p,0)\leq  \mu_\alpha^{max} <\mu_\alpha.
\end{equation*}

So we have $$\Delta_{\mu_\alpha}(p,0)<0.$$
 Hence by Lemma
\ref{delta=0} the operator $h_\alpha(p),\,p \in \T^3$ has a unique
negative eigenvalue.

 $(ii)$ Let $\mu_\alpha^{max}\geq \mu_\alpha>\mu_\alpha^0.$
 By Hypothesis \ref{hypothD}  for all $p \in
\T^3,\,p\neq 0$ the inequality
\begin{equation*}
\Delta_{\mu_\alpha^0}(p,0)=1-\mu_\alpha^0\Lambda_\alpha(p,0)>0
\end{equation*}
holds, that is,
\begin{equation*}
 \Lambda_\alpha^{-1}(p,0)>\mu_\alpha^0.
\end{equation*}

Let us introduce the notation
\begin{equation}\label{ddd}
D_{\mu_\alpha}\equiv\{p\in \T^3:
\Lambda_\alpha^{-1}(p,0)<\mu_\alpha\}.
\end{equation}

Since the function $\Lambda^{-1}_\alpha(\cdot,0)$ is the continuous
on $\T^3$ and $\Lambda_\alpha^{-1}(0,0)=\mu_\alpha^0$ we have that
$D_{\mu_\alpha}\neq \T^3$ is a non void open set.

Thus we have
\begin{equation*}
\Delta_{\mu_\alpha}(p,0)<0\quad\mbox{for all}\quad p \in
D_{\mu_\alpha}.
\end{equation*}

Hence by Lemma \ref{delta=0} the operator $h_\alpha(p),\,p \in
D_{\mu_\alpha}$ has a  unique negative eigenvalue.

For all $ p \in \T^3\setminus D_{\mu_\alpha}$ we have
\begin{equation}\label{inDmu}
 \Lambda_\alpha^{-1}(p,0)\geq
\mu_\alpha,\quad \mbox{ that is,}\quad
 \Delta_{\mu_\alpha}(p,0) \geq 0.
\end{equation}

For any $p\in {\T}^3$ the function $\Delta_{\mu_\alpha}(p,z)$ is
decreasing  in $z\in (-\infty,m_\alpha(p))$ and\\ $\lim_{z\to
-\infty} \Delta_{\mu_\alpha}(p,z)=1$, hence from \eqref{inDmu} for
any $z<0$ and $p \in \T^3\setminus D_{\mu_\alpha}$
 we have $\Delta_{\mu_\alpha}(p,z)>
\Delta_{\mu_\alpha}(p,0)\geq 0$.
 Then by Lemma \ref{delta=0}
 for $p \in \T^3\setminus D_{\mu_\alpha}$  the operator
$h_\alpha(p)$ has no negative eigenvalues.

$(iii)$ Let $0<\mu_\alpha\leq \mu_\alpha^0.$ Then by Hypotheses
\ref{hypothD} we have
\begin{equation*}
\Delta_{\mu_\alpha}(p,0)>0\quad \mbox{for all}\quad p\in \T^3.
\end{equation*}

For any $p\in {\T}^3$ the function $\Delta_{\mu_\alpha}(p,z)$ is
decreasing  in $z\in (-\infty,m_\alpha(p))$ and $$\lim_{z\to
-\infty} \Delta_{\mu_\alpha}(p,z)=1$$ and
 we have $\Delta_{\mu_\alpha}(p,z)>
\Delta_{\mu_\alpha}(p,0),\,z<0.$ Then by Lemma \ref{delta=0} for
all $p\in \T^3$ the operator $h_\alpha(p)$ has no  negative
eigenvalues.
\end{proof}

The following decomposition plays a crucial role  for the proof of
the asymptotics \eqref{asym.K}.
 \begin{lemma}\label{razlojeniya} Assume Hypothesis \ref{hypoth u}
 and \ref{hyp.varphi} are
fulfilled.
 Then for any $p\in U_{\delta}(0),\delta>0$ sufficiently small, and
 $z\leq 0$  the following decomposition holds:
\begin{align}\label{raz}
&\Delta_{\mu_\alpha}(p,z)=
 \Delta_{\mu_\alpha}(0,0)+\\
 &\hspace{1cm} \frac{4\sqrt{2}\pi^2 \mu_\alpha \varphi^2_\alpha(0)}
 {l_\beta^{\frac{3}{2}} \det(U)^{\frac{1}{2}}
 }\sqrt{m_\alpha(p)-z}+
\Delta^{(02)}_{\mu_\alpha}(m_\alpha(p)-z)+
\Delta^{(20)}_{\mu_\alpha}(p,z),\nonumber
\end{align}
where $\Delta^{(02)}_{\mu_\alpha}(m_\alpha(p)-z)$
 (resp. $ \Delta^{(20)}_{\mu_\alpha}(p,z)$) is a function behaving like
 $O({m_\alpha(p)-z})$ (resp. $O(|p|^2)$)  as $|{m_\alpha(p)-z}| \to 0 $ (resp.
 $p\to 0$).
\end{lemma}
\begin{proof}
 Let
 \begin{equation}\label{U-alpha.1}
 U_\alpha(p,q)=u^{(\alpha)}_p
(q+q_\alpha(p))-m_{\alpha}(p),
\end{equation}
where $q_\alpha(p)\in\T^3$ is an analytic function in $p\in
U_{\delta}(0)$ (see Lemma \ref{minimum}) and   is the non-degenerate
minimum point of the function $u^{(\alpha)}_p (q)$ for any $p \in
U_{\delta}(0).$

We define the function $\tilde\Delta_{\mu_\alpha}(p,w)$ on $
U_\delta(0)\times {\C}_{+}$  by
 $$
\tilde\Delta_{\mu_\alpha}(p,w)=\Delta_{\mu_\alpha}(p,m_\alpha(p)-w^2),$$
where ${\C}_{+}=\{z\in \C: \Re z>0\}$. Using \eqref{U-alpha.1} the
function $\tilde\Delta_{\mu_\alpha}(p,w)$ is represented as
\begin{equation*}
\tilde {\Delta}_{\mu_\alpha}( p, w) = 1-\mu_\alpha
\int\limits_{{\T}^3}
\frac{\varphi^2_{\alpha}(q+q_\alpha(p))}{U_\alpha(p,q)+w^2 } d q .
\end{equation*}

Let $V_{\delta}(0)$ be a complex $\delta$-neighborhood of the point
$w=0 \in \C$. Denote by $\Delta^{*}_{\mu_\alpha}(p,w)$ the analytic
continuation of the function $\tilde\Delta_{\mu_\alpha}(p,w)$ to the
region $U_\delta(0)\times ({\C}_{+} \cup V_{\delta}(0))$. Since the
functions $|\varphi_\alpha(q)|$ and $U_\alpha(p,q)$
 are even
(see Lemma \ref{ras.u(p,q)})
  we have that
$\Delta^{*}_{\mu_\alpha}(p,w)$
  is even in $p\in U_\delta(0).$

Therefore
\begin{equation}
\label{raz1}
 \Delta^{*}_{\mu_\alpha}(p,w)=\Delta^{*}_{\mu_\alpha}(0,w)+
 \tilde\Delta^{(20)}_{\mu_\alpha}(p,w),
\end{equation}
 where $\tilde\Delta^{(20)}_{\mu_\alpha}(p,w)=O(|p|^2)$
 uniformly in  $w \in {\C_{+}}$ as $p\to 0$ (see also \cite{Ltmf92}).
A Taylor series expansion gives
 \begin{equation}
\label{raz2}
\Delta^{*}_{\mu_\alpha}(0,w)=\Delta^{*}_{\mu_\alpha}(0,0)+
\tilde\triangle^{(01)}_{\mu_\alpha}(0,0)w+
\tilde\triangle^{(02)}_{\mu_\alpha}(0,w)w^2,
\end{equation} where
$\tilde\triangle^{(02)}_{\mu_\alpha}(0,w)=O(1)\quad\text{as}\quad
w\to 0.$

 A simple computation shows that
\begin{equation}\label{partial}
\frac{\partial \Delta^{*}_{\mu_\alpha}(0,0)}{\partial
w}=\tilde\triangle^{(01)}_{\mu_\alpha}(0,0)= \frac{ 4\sqrt{2}\pi^2
\mu_{\alpha} \varphi_{\alpha}^2(0)}{l_\beta^ {\frac{3}{2}}
\det(U)^{\frac{1}{2}} }.\end{equation}

The representations \eqref{raz1} , \eqref{raz2} and the equality
\eqref{partial} give \eqref{raz}.
\end{proof}

\begin{corollary}\label{razl.lemma.natijasi.}
 Assume Hypothesis
\ref{hypoth u} and \ref{hyp.varphi} are fulfilled and let
$h_\alpha(0)$ have a zero energy resonance. Then there exists a
number
 $\delta>0$  so that
for any $p\in U_{\delta}(0) $ and
 $z \leq 0$ the following decomposition holds
\begin{align}\label{raz.cor2}
\Delta_{\mu^0_\alpha}(p,z)=\frac{4\sqrt{2}\pi^2 \mu^0_\alpha
\varphi^2_\alpha(0)}
 {l_\beta^{\frac{3}{2}} \det(U)^{\frac{1}{2}}}\sqrt{m_\alpha(p)-z}+
\Delta^{(02)}_{\mu^0_\alpha}({m_\alpha(p)-z})+
\Delta^{(20)}_{\mu^0_\alpha}(p,z),
\end{align}
where the functions $\Delta^{(02)}_{\mu_\alpha}({m_\alpha(p)-z})$
and $\Delta^{(20)}_{\mu_\alpha}(p,z)$ are the same as in Lemma
\ref{razlojeniya}.
\end{corollary}
\begin{proof}
The proof of Corollary \ref{razl.lemma.natijasi.} immediately
follows from decomposition \eqref{raz} and the equality
$\Delta_{\mu^0_\alpha}(0,0)=0.$
\end{proof}

\begin{corollary}\label{razl.lemma.natijasi.0}
 Assume Hypothesis
\ref{hypoth u} and \ref{hyp.varphi} are fulfilled and
$\varphi_\alpha(0)\neq 0.$
 Then there
exists $\delta>0$  such that for any $p\in U_{\delta}(0),\,p\neq 0 $
\begin{equation}\label{Lamb.ineq}
\Delta_{\mu_\alpha^0}(p,0)>0,\quad {that\,\, is,}\quad
\Lambda_\alpha(p)<\Lambda_\alpha(0),
\end{equation}
where $\Lambda_\alpha(p)$ is defined using the integral
\eqref{Lamb}.
\end{corollary}
\begin{proof}
Let $\varphi_\alpha(0)\neq 0$, then by  Corollary
\ref{razl.lemma.natijasi.} and the asymptotics (see \eqref{min.raz})
\begin{equation}\label{malfa}
m_\alpha(p)= \frac{l_1l_2-l^2}{2l_\alpha}(Up,p)+O(p^4)\quad
\mbox{as} \quad p\to 0
\end{equation}  we get $$ \frac{4\sqrt{2}\pi^2
\mu^0_\alpha \varphi^2_\alpha(0)}
 {l_\beta^{\frac{3}{2}} \det(U)^{\frac{1}{2}}}\sqrt{m_\alpha(p)}>
|\Delta^{(20)}_{\mu^0_\alpha}(0,p)| $$ for $p\in U_\delta(0),p\neq
0,\,\delta$ sufficiently small. This inequality gives
\eqref{Lamb.ineq}.
\end{proof}

\begin{lemma}\label{D.ineq}
 Assume Hypothesis
\ref{hypoth u}, \ref{hyp.varphi} and \ref{hypothD} are fulfilled
and the operator $h_\alpha(0)$ has a zero-energy resonance. Then
there exist positive numbers $c,C$ and $\delta$ such that
\begin{equation}\label{c<(.,.)<c}
c |p| \leq \Delta_{\mu^0_\alpha}(p,0 ) \leq C |p|\quad\mbox{for
any}\quad p\in U_\delta(0)
 \end{equation}
and
\begin{equation}\label{(.,.)>c}
\Delta_{\mu^0_\alpha}(p,0 ) \geq c  \quad\mbox{for any}\quad\mbox
p\in \T^3\setminus U_\delta(0).
 \end{equation}
\end{lemma}
\begin{proof}
From \eqref{raz.cor2} and \eqref{malfa} we get \eqref{c<(.,.)<c} for
some positive numbers $c,C$.

By Hypothesis \ref{hypothD}  we get
$\Delta_{\mu^0_\alpha}(p,0)>0,\, p\neq 0$. Since
$\Delta_{\mu^0_\alpha}(p,0 )$ is continuous on $\T^3$ and
$\Delta_{\mu^0_\alpha}(0,0 )=0$ we have \eqref{(.,.)>c} for some
$c>0.$
\end{proof}
 \begin{lemma}\label{main.ineq}
 Assume Hypothesis
\ref{hypoth u} and \ref{hypothD} are fulfilled and let the operator
$h_\alpha (0)$ have a zero eigenvalue,
 then there exist  numbers $\delta>0$ and $c>0$ so that
 the following inequalities hold
\begin{align*}
|\Delta_{\mu^0_\alpha}(p,0)|\geq c p^2 \quad \mbox{for all}\quad
p\in U_\delta(0),\\ |\Delta_{\mu_\alpha}(p,0)|\geq c
 \quad \mbox{for all}\quad
p\in \T^3\setminus U_\delta(0).
\end{align*}
\end{lemma}
\begin{proof} Let the operator $h_\alpha(0)$ have a zero
eigenvalue. By Lemma \ref{resonance} we have
$\mu_\alpha={\mu^0_\alpha}$, $\varphi_\alpha(0)=0$ and
$\Delta_{\mu^0_\alpha}(0,0)=0$. By Hypothesis \ref{hypothD} the
function $\Delta_{\mu^0_\alpha}(p,0)=1-\mu^0_\alpha
\Lambda_{\alpha}(p)$ has a unique non-degenerate minimum at $p=0$.
Then there exist positive numbers $\delta$ and $c$ such that the
statement of the lemma is fulfilled.
\end{proof}

\section{The essential spectrum of the  operator $H$}

In this section we introduce a multiplication operator perturbed
by a partial integral operator and  prove Theorem \ref{ess}.

We consider the operator $H_\alpha$ acting on the Hilbert space
$L_2(({\T}^3)^2)$ as $$
  H_\alpha=H_0-\mu_\alpha V_\alpha,\,\,\alpha=1,2.
$$

The operator $H_1$ (resp. $H_2$) commutes with all multiplication
operators by functions $w_1(q)$ (resp.  $w_2(p)$) on $L_2 ((
{\T}^3 )^2).$

 Therefore the  decomposition of the space $L_2 (( {\T}^3 ) ^2)
$ into the direct integral
 $$L_2 (( {\T}^3 ) ^2) = \int\limits_{p\in {\T}^3}
\oplus L_2({\T}^3) d p $$ yields for the operator $H_{\alpha}$ the
decomposition into the direct integral
 \begin{equation}\label{decompose} H_{\alpha} =  \int\limits_{p\in {\T}^3}
 \oplus h_{\alpha}(p) dp.
 \end{equation}

Here the fiber operators $h_\alpha(p),\,p\in \T^3,$ are defined by
\eqref{h_alpha}.
\subsection{The spectrum of the operators $H_\alpha$  }
The representation \eqref{decompose} of the operator $H_\alpha$ and
the theorem on decomposable operators (see \cite{ReSiIV}) imply the
following lemma:

\begin{lemma}\label{spec}
For the spectrum  $\sigma(H_\alpha)$  of the $H_\alpha$ the
equality
\begin{equation*}
\sigma(H_\alpha)= \cup_{p\in {\T}^3} \sigma_d(h_\alpha(p))\cup [0,
M]
\end{equation*}
holds.
\end{lemma}
\qed

Set
\begin{align}\label{two.branch}
&\sigma_{two}(H_\alpha)=\cup_{p\in {\T}^3} \sigma_d(h_\alpha(p)).
\end{align}
Now we precisely describe the location and structure of the
spectrum of $H_\alpha$.
\begin{lemma}\label{two.bran.lem}
Assume Hypotheses \ref{hypoth u}
is fulfilled.\\
 (i) Let $\mu_\alpha >\mu_\alpha^{max}$, then
\begin{equation*}
\sigma(H_\alpha)= [a_\alpha,b_\alpha]\cup [0, M]\quad
\mbox{and}\quad b_\alpha<0.
\end{equation*}
(ii) Let $ \mu_\alpha^{max}\geq  \mu_\alpha> \mu_\alpha^0$, then
\begin{equation*}
\sigma(H_\alpha)= [a_\alpha, M]\quad \mbox{and}\quad a_\alpha<0.
\end{equation*}
(iii) Let $   \mu_\alpha^0\geq \mu_\alpha >0$, then
\begin{equation*}
\sigma(H_\alpha)= [0,M].
\end{equation*}
\end{lemma}
\begin{proof}
$(i)$ Let $\mu_\alpha >\mu_\alpha^{max}$. Then by Lemma \ref{aaa}
for all $p\in \T^3$ the operator $h_\alpha(p),\,\alpha=1,2,$ has a
unique  negative  eigenvalue $z_\alpha(p)<m_\alpha(p)$.

By Hypotheses \ref{hypoth u} and \ref{hyp.varphi} and Lemma
\ref{delta=0} $z_\alpha: p\in  \T^3 \to  z_\alpha(p)$ is a real
analytic function on $\T^3.$

Therefore $\Im z_\alpha$ is a connected  closed subset of
$(-\infty,0)$, that is, $\Im z_\alpha=[a_\alpha,b_\alpha]$ and
$b_\alpha<0$ hence $\sigma_{two}(H_\alpha)=[a,b].$

$(ii)$ Let $\mu_\alpha^{max}\geq \mu_\alpha >\mu_\alpha^0$. Then
by Lemma \ref{aaa} there exists a non void  open set
$D_{\mu_\alpha}$ (see \eqref{ddd}) such that for any $p\in
D_{\mu_\alpha} $ the operator $h_\alpha(p)$ has a unique negative
eigenvalue $z_\alpha(p)$.

Since for any $p\in \T^3$ the operator $h_\alpha(p)$ is bounded
and $\T^3$ is compact set there exist positive number $C$ such
that $sup_{p\in \T^3}||h_\alpha(p)||\le C$ and for any $p\in \T^3$
we have
\begin{equation*}
\sigma(h_\alpha(p))\subset [-C,C].
\end{equation*}

For any $q\in \partial D_{\mu_\alpha}=\{p\in \T^3:
\Delta_{\mu_\alpha}(p,0)=0\}$ there exist $\{p_n\}\subset
D_{\mu_\alpha}$ such that $p_n\to q$ as $n\to \infty.$ Set
$z_n=z(p_n).$ By Lemma \ref{aaa}  we have $\{z_n\}\subset [-C,0]$
and without loss of a generality one may assume that $z_{n}\to z_0$
as $n\to \infty.$

The function $\Delta_{\mu_\alpha}(p,z)$ is continuous in $\T^3\times
(-\infty,0]$ and hence
$$
0=\lim\limits_{n\to \infty}\Delta(p_{n},z_{n}) =\Delta(q,z_0).
$$

Since for any $p\in \T^3$ the function
$\Delta_{\mu_\alpha}(p,\cdot)$ is decreasing  in $(-\infty,m_w(p)]$
and $p\in
\partial D_{\mu_\alpha}$ we can see that $\Delta_{\mu_\alpha}(p,z_0)=0$ if and only if
$z_0=0.$

For any $p\in \partial D_{\mu_\alpha}$ we define
$$
z_\alpha(p)=\lim\limits_{q\to p,\,q\in D_{\mu_\alpha}}z_\alpha(q)=0.
$$
Since the function $z_\alpha(p)$ is continuous on the compact set
$D_{\mu_\alpha}\cup\partial D_{\mu_\alpha}$ and
$z_\alpha(p)=0,\,\,p\in
\partial D_{\mu_\alpha}$ we conclude that
$$
\Im z_\alpha=[a_\alpha,0],\quad a_\alpha<0.
$$

Hence the set
$$
\{z \in \sigma_{two}(H_\alpha): z\le0\}=
 \cup_{p\in \T^3}
\sigma_d(h_\alpha(p))\cap (-\infty,0]
$$
coincides with the set $\Im z_\alpha=[a_\alpha,0].$

Then Lemma \ref{spec} and \eqref{two.branch} complete the proof of
$(ii).$

$(iii)$ Let $\mu_\alpha^0\geq \mu_\alpha >0$. Then by Lemma
\ref{aaa} for all $p\in \T^3$ the operator $h_\alpha(p)$ has no
negative eigenvalue.

Hence we have
\begin{equation*}
 \sigma(H_\alpha)=[0,M].
\end{equation*}
\end{proof}

\subsection{The Faddeev type integral equation} In this section we derive a Faddeev type system of
integral equations and prove an analogue of the Birman-Schwinger
principle.

Denote by $R_\alpha(z),\,\alpha=1,2,$ resp. $R_0(z)$ the resolvent
operator of $H_\alpha,\,\alpha=1,2,$ resp. $H_0$.

Let $W_\alpha(z),\alpha=1,2,$ be the operators on
$L_2(({\T}^3)^2)$ defined as
 \begin{equation*}\no
W_\alpha(z)={I}+\mu_{\alpha}V_\alpha^{\frac{1}{2}}
R_\alpha(z)V_\alpha^{\frac{1}{2}},\quad z\in \rho(H_{\alpha}),
 \end{equation*}
where $ I$ is the identity operator on $L_2((\T^3)^2)$ and
 $\rho(H_{\alpha})=\C\setminus \sigma(H_\alpha)$ is the
 set of all regular points of the operator $H_{\alpha}$.

 One can check that
 \begin{equation}\no
W_\alpha(z)=( I-\mu_{\alpha}V^{\frac{1}{2}}_\alpha
R_0(z)V^{\frac{1}{2}}_\alpha)^{-1},\,\alpha=1,2.
 \end{equation}

Let  $ {\bf A}(z),\, z \in {\C} \setminus
(\sigma(H_1)\cup\sigma(H_2))$ be the operator on
$L^{(2)}_2(({\T}^3)^2)$ with  the entries
\begin{align*}\no
& {\bf A}_{ \alpha\alpha } (  z) =0, \\
 &{\bf A}_{ \alpha\beta } ( z) =\sqrt{\mu_{\alpha}
\mu_{\beta}
 }W_{\alpha}(z)V_\alpha^{\frac{1}{2}}
R_0(z)V_\beta^{\frac{1}{2}},\quad \alpha\neq
\beta,\,\,\alpha,\beta=1,2.
\end{align*}

The following theorem is an analogue of the  result
 for the three-particle discrete Schr\"{o}dinger operators
with the zero-range interactions and may be proven in a way similar
to the one in \cite{Lfa93}.
\begin{theorem}\label{rrr}
 The number $z \in {\C} \setminus
(\sigma(H_1)\cup\sigma(H_2))$ is an eigenvalue of the operator $H$
if only if the number 1 is an eigenvalue of ${\bf A} (z)$. Moreover
the eigenvalues $z$ and $1$ have the same multiplicities.
\end{theorem}
\begin{proof}
 Let $z \in \C \setminus (\sigma(H_1)\cup\sigma(H_2))$ be
the eigenvalue of  $H,$
 that is,  the equation
\begin{equation}\label{AZ1}
f=R_0(z)\sum_{\alpha=1}^2 \mu_{\alpha} V_\alpha f,\quad f \in
L_2((\T^3)^2)
 \end{equation}
has a nontrivial solutions.

 Multiplying  \eqref{AZ1} from the left by
 $\sqrt{\mu_{\alpha}}V^{\frac{1}{2}}_{\alpha}$ and setting
  $\varphi_\alpha=\sqrt{\mu_{\alpha}} V^{\frac{1}{2}}_{\alpha}f$
 we have the following system of two equations
\begin{align}\label{AZ2}
\varphi_\alpha=\sum_{\beta=1}^{2}
\sqrt{\mu_{\alpha}\mu_{\beta}}V^{\frac{1}{2}}_{\alpha}R_0(z)
V^{\frac{1}{2}}_{\beta} \varphi_\beta, \quad \alpha=1,2,\\
\varphi=(\varphi_1,\varphi_2)\in L^{(2)}_2(({\T}^3)^2)\no
\end{align}
and  this system of equations has a nontrivial solution if and
only if the system of equations \eqref{AZ1} has a nontrivial
solution and
 the linear subspaces of solutions of
\eqref{AZ1} and \eqref{AZ2} have the same dimensions.

Since the operator
 $
W_\alpha(z)=({I}-{\mu_{\alpha}}V^{\frac{1}{2}}_\alpha
R_0(z)V^{\frac{1}{2}}_\alpha)^{-1},$ where $I$ is identity operator
on $L_2((\T^3)^2),$  is invertible the system of equations
\eqref{AZ2} is equivalent to the following system of two equations
\begin{align*}
&\varphi_\alpha= \sqrt{\mu_{\alpha}\mu_{\beta}}
W_{\alpha}(z)V^{\frac{1}{2}}_{\alpha}R_0(z)V^{\frac{1}{2}}_{\beta}
\varphi_\beta, \quad \alpha\neq \beta,\,\alpha,\beta=1,2,
\end{align*}
that is,
\begin{equation}\no
 {\bf A}(  z) \varphi = \varphi,\quad  \varphi=(\varphi_1,\varphi_2)
 \in
 L^{(2)}_2(({\T}^3)^2).
\end{equation}
\end{proof}

 Let ${\bf T} (z),  z \in \C
  \setminus (\sigma(H_1)\cup\sigma(H_2))$ be the operator on  $L^{(2)}_2({\T}^3) $ with the entries
\begin{align*}
 &{\bf T}_{11} (z) = {\bf T}_{ 22}(z) = 0, \\
 &({\bf T}_{12}(z)w)(q)= \frac{\sqrt{\mu_1\mu_2}}{\Delta_{\mu_1}(q,z)}
\int \limits_{{\T}^3} \frac{\varphi_1(t)\varphi_2(q)}{
u(t,q)-z}w(t)dt,\nonumber\\
 &({\bf
T}_{21}(z)w)(p)=\frac{\sqrt{\mu_1\mu_2}}{\Delta_{\mu_2}(p,z)} \int
\limits_{{\T}^3} \frac{\varphi_1(p) \varphi_2(t)}{
u(p,t)-z}w(t)dt.\nonumber
\end{align*}

Let $ \Phi=\text{diag}\{\Phi_1,\Phi_2\}:L^{(2)}_ 2(({\T}^3)^2)\to
L^{(2)}_ 2({\T}^3) $  be the operator with the entries
\begin{align}\label{izom}
&(\Phi_1 f)(q)= \frac{1}{||\varphi_1||}\int_{{\T}^3}\varphi_1(t)
f(t,q)dt,\quad (\Phi_2
f)(p)=\frac{1}{||\varphi_2||}\int_{{\T}^3}\varphi_2(t) f(p,t)dt,\\
&\hspace{4cm} f\in L_2((\T^3)^2)\nonumber
\end{align}
 and
$\Phi^*=\text{diag}\{\Phi^*_1,\Phi^*_2\}$ its adjoint.

\begin{proposition}\label{propos}
Let $T_1,T_2$ be bounded operators. If $z\neq0$ is an eigenvalue
of $T_1T_2$ then $z$ is an eigenvalue for $T_2T_1$ as well with
the same algebraic and geometric multiplicities.
\end{proposition}
 This proposition is well known and its proof is omitted.

\begin{lemma}\label{fadd}
For each  $z \in \C \setminus (\sigma(H_1)\cup\sigma(H_2))$ the
following equality
 \begin{equation}\label{AZTZ}
{\bf A}(z)=\Phi^*{\bf T}(z)\Phi
 \end{equation}
 holds and the nonzero eigenvalues of  the operators ${\bf A}(z)$ and ${\bf
T}(z)$ coincide and have the same  algebraic and geometric
multiplicities.
\end{lemma}
\begin{proof}
One can easily check that the equalities
\begin{equation}\label{faddeev}
 (V^{1/2}_1 f)(p,q)=\varphi_1(p)(\Phi_1 f)(q),\,
 (V^{1/2}_2 f)(p,q)=\varphi_2(q)(\Phi_2 f)(p)
\end{equation} hold.
The equalities \eqref{faddeev}
  imply the
 equality  \eqref{AZTZ}.

By virtue of Proposition \ref{propos} and \eqref{AZTZ} the nonzero
eigenvalues of the operators ${\bf A}(z)$ and ${\bf T}(z)$ are
same.
\end{proof}

 Now we establish the location of the essential
spectrum of $H.$ For any  $z \in {\C} \setminus
(\sigma(H_1)\cup\sigma(H_2)) $ the kernels of the operators ${\bf
T}_{\alpha\beta}(z),\alpha,\beta=1,2,$ are continuous functions on
$({\T}^3)^2$. Therefore the Fredholm determinant $\det\big ({\bf
I}-{\bf T}( z) \big )$ of the operator ${\bf I}-{\bf T}( z)$,
where ${\bf I}$ is the identity operator on $L^{(2)}_2({\T}^3),$
exists and is a real-analytic function on ${\C} \setminus
(\sigma(H_1)\cup\sigma(H_2)).$
 According to the Fredholm theorem the following lemma
holds.
\begin{lemma}\label{nol} The number $z \in {\C} \setminus (\sigma(H_1)\cup\sigma(H_1))$
is an eigenvalue of the operator $H$
 if and only if
$$ \det\big ({\bf I}-{\bf T}( z) \big )=0. $$
\end{lemma} \qed

\begin{theorem}\label{ess1}
For the essential spectrum of the operator $H$ the inequality
 $$
\sigma_{ess}(H)=\cup_{\alpha=1}^2 \cup_{p\in
{\T}^3}\sigma_d(h_\alpha(p)) \cup [0,M]
$$
holds.
\end{theorem}

\begin{proof} By the definition of the essential spectrum, it is
easy to show that $ \sigma(H_1)\cup\sigma(H_1) \subset
{\sigma}_{ess}(H).$ Since the function $\det\big ({\bf I}-{\bf T}(
z) \big )$ is analytic in ${\C}\setminus
(\sigma(H_1)\cup\sigma(H_1))$ by Lemma \ref{nol} we conclude that
the set
\begin{equation*}
\sigma(H)\setminus (\sigma(H_1)\cup\sigma(H_1))=\{z \in
\C\setminus(\sigma(H_1)\cup\sigma(H_1)):\, \det\big ({\bf I}-{\bf
T}( z) \big )=0\}
\end{equation*}
is discrete. Thus
  $$
  \sigma(H)\setminus (\sigma(H_1)\cup\sigma(H_1))
  \subset  \sigma(H)\setminus{\sigma}_{ess}(H).
  $$
Therefore the inclusion ${\sigma}_{ess}(H)\subset
(\sigma(H_1)\cup\sigma(H_1))$ holds.
\end{proof}
{\bf Proof of Theorem \ref{ess}}. The proof of Theorem \ref{ess}
follows from Theorem \ref{ess1} and Lemma \ref{two.branch}.

\begin{theorem}\label{upper.eige}
 The
operator $H=H_0-V$ has no eigenvalue lying on the right hand side
of the essential spectrum $\sigma_{ess}(H).$
\end{theorem}
\begin{proof}
Since $V=\mu_1V_1+\mu_2V_2$ is a positive operator we have that
the operator $H=H_0-V$ has no eigenvalues larger than $M$.
\end{proof}

\subsection{Birman-Schwinger principle}

We recall that $ \tau_{ess}(H)$ denotes the bottom of the essential
spectrum
 and $N(z)$ the number of eigenvalues of $H$ lying below
 $z \leq \tau_{ess}(H).$

Let  $ A(z),\, z <\tau_{ess}(H)$ be the operator on
$L^{(2)}_2(({\T}^3)^2)$ with  the entries
\begin{align*}\no
& {A}_{ \alpha\alpha } (  z) =0, \\
 &{A}_{ \alpha\beta } ( z) =\sqrt{\mu_1\mu_2}\,
W^{\frac{1}{2}}_{\alpha}(z)V_\alpha^{\frac{1}{2}}
R_0(z)V_\beta^{\frac{1}{2}}W^{\frac{1}{2}}_{\beta}(z) ,\quad
\alpha\neq \beta,\,\,\alpha,\beta=1,2.
\end{align*}

 For a bounded self-adjoint operator $B,$ we define
$n(\lambda,B)$ as
$$
n(\lambda,B)=sup\{ dim F: (Bu,u) > \lambda,\, u\in F,\,||u||=1\}.
$$
$n(\lambda,B)$ is equal to the infinity if $\lambda$ is in
essential spectrum of $B$ and if $n(\lambda,B)$ is finite, it is
equal to the number of the eigenvalues of $B$ bigger than
$\lambda$.
 By the definition of $N(z)$ we have $$
N(z)=n(-z,-H),\,-z > -\tau_{ess}(H). $$ The following lemma is a
realization of well known Birman-Schwinger principle for the
operator $H$ (see \cite{ALzM,Sob,Tmsj94} ).
\begin{lemma}\label{b-s}
The operator ${A}(z)$ is compact and continuous in $z<\tau_{ess}(H)$
and
$$ N(z)=n(1,A(z)). $$
\end{lemma}
\begin{proof}
This Lemma is deduced by the arguments of \cite{Sob}. Since
$H=H_0+V,$ and $H_0-zI$ is positive and invertible for
$z<\tau_{ess}(H),$ one has $u\in L_2(({\T}^3)^2)$ and
 $((H-zI)u,u)<0$ if and only if
$(R^{\frac{1}{2}}_0(z)V R^{\frac{1}{2}}_0(z)-I)v,v)>0$ and
$v=(H_0-z)^{\frac{1}{2}}u.$

It follows that
$N(z)=n(1,R^{\frac{1}{2}}_0(z)VR^{\frac{1}{2}}_0(z)).$ We have the
following decomposition:
$$
R_{0}^{\frac{1}{2}}(z)\,V\,
  R_{0}^{\frac{1}{2}}(z))=Z^{*}Z,
  $$
with $Z:L_2((\T^3)^2)\rightarrow L^{(2)}_2((\T^3)^2)$ defined by
$$
Z=( \sqrt{\mu_{1}}R_{0}^{\frac{1}{2}}(z)V_1^{\frac{1}{2}},
\sqrt{\mu_{2}}R_{0}^{\frac{1}{2}}(z)V_2^{\frac{1}{2}}).
$$
Then by Proposition \ref{propos} we get
$$
n(\lambda,Z^*Z)=n(1,ZZ^*).
$$
Consequently,
$$
N(z)=n(1,ZZ^*).
$$
Since $z< \tau_{ess}(H)$ the operator
$I-{\mu_\alpha}V^{\frac{1}{2}}_\alpha
R_0(z)V^{\frac{1}{2}}_\alpha,\, \alpha=1,2,$ is invertible, and
$$
(I-{\mu_\alpha}V^{\frac{1}{2}}_\alpha
R_0(z)V^{\frac{1}{2}}_\alpha))^{-\frac{1}{2}}=W_\alpha^{\frac{1}{2}}(z)>0,
\alpha=1,2.
$$

Let  $\mathbb{I}$ be the identity operator on
$L^{(2)}_2((\T^3)^2).$ A direct calculations shows that $y\in
L^{(2)}_2((\T^3)^2)$ and $((ZZ^*-\mathbb{I})y,y)>0$ if and only if
$ ((A(z)-\mathbb{I})f,f)>0$ and
$y_\alpha=(I-{\mu_\alpha}V^{\frac{1}{2}}_\alpha
R_0(z)V^{\frac{1}{2}}_\alpha)^{\frac{1}{2}}f_\alpha,\,\alpha=1,2,$
and that $n(1,ZZ^*)=n(1,A(z)).$
\end{proof}

In our analysis of the spectrum of $H$ the crucial role is played by
the compact integral operator  $ T( z),\, z <\tau_{ess}(H)$ in the
space $L^{(2)}_2({\T}^3)$ with the entries

\begin{align*}
 &T_{11 } (  z) = T_{22}( z) = 0, \\
 &(T_{12}(z)\omega)(q)= \sqrt {\mu_1\mu_2}
\int \limits_{{\T}^3} \frac{\varphi_1(t)
\varphi_2(q)}{\sqrt{\Delta_{\mu_1}(q,z)}
\sqrt{\Delta_{\mu_2}(t,z)}(u(t,q)-z)}\omega(t)dt,\\
&(T_{21}(z)\omega)(p)=\sqrt{\mu_1 \mu_2} \int \limits_{{\T}^3}
\frac{\varphi_1(p) \varphi_2(t)}{\sqrt{\Delta_{\mu_2}(p,z)}
\sqrt{\Delta_{\mu_1}(t,z)}(u(p,t)-z)}\omega(t)dt,\,\quad w\in
L_2(\T^3).
\end{align*}

Using the equality \eqref{faddeev} one may verify that $$
{A}(z)=\Phi^*T(z)\Phi, $$ where the operator $\Phi$ is defined in
\eqref{izom}.

 Since the operator $\Phi^*T(z)$
resp. $\Phi$ is a bounded in $L^{(2)}_2({\T}^3)$ resp.
$L^{(2)}_2(({\T}^3)^3)$, by Proposition \ref{propos} and the
equality $\Phi\Phi^*T(z)=T(z)$ we have $n(1,A(z))=n(1,{T}(z)).$

\section{ The finiteness of the  number of eigenvalues of the  operator $H$.}

We start the proof of $(i)$ of Theorem \ref{fin} with one elementary
lemma.
\begin {lemma}\label{G-S} Assume Hypotheses \ref{hypoth u},
\ref{hyp.varphi} and \ref{hypothD} are fulfilled and let
$\mu_\alpha =\mu_\alpha^0$ for all $\alpha=1,2,$ and either
$\varphi_1(0)=0$ , $\varphi_2(0)\neq 0$ or $\varphi_1(0)\neq 0$ ,
$\varphi_2(0)= 0$ or $\varphi_1(0)=0$ , $\varphi_2(0)= 0$.
 Then  the operator $T(z)$ belongs to the Hilbert-Schmidt class and
is continuous from the left up to $z=0$.
\end{lemma}
\begin{proof}
We prove  Lemma \ref{G-S}
 in the case $\mu_\alpha =\mu_\alpha^0,\alpha=1,2,$
  and $\varphi_1(0)=0$, $\varphi_2(0)\neq
 0$ ( the other cases are handled in a similar way).

Since the function $\varphi_1(p)$ is analytic and $\varphi_1(0)=0$
we have $|\varphi_1(p)|\leq C|p|$ for some $C>0$.
 By virtue of Lemmas \ref{D.ineq}, \ref{main.ineq}  and
\ref{U.ineq} the kernel of the operator $T_{12}(z),\,z\leq 0$ is
estimated by
$$ C\Big (
 \frac{\chi_{\delta}(p)}{|p|}+1)(
\frac{|q| \chi_{\delta}(p)\chi_{\delta}(q)}{p^2+q^2}+1
 )(\frac{\chi_{\delta}(q)}{|q|^{\frac{1}{2}}}+1)
 \Big),
$$ where  $ \chi_{\delta}(p)$ is the characteristic function of
$U_\delta(0).$

 Since the latter function is square integrable on $(\T^3)^2$ we
have  that the operators $T_{12}(z)$ and $T_{21}(z)=T^*_{12}(z)$
are Hilbert-Schmidt operator.

The kernel function of $T_{\alpha\beta}(z)$ is continuous in $p,q
\in \T^3,\,z<0$ and square integrable
 on $(\T^3)^2$ as $z\leq 0$.
Then by the dominated convergence theorem the operator
$T_{\alpha\beta}(z)$ is continuous from the left up to $z=0.$
\end{proof}

We are now ready for the

 {\bf Proof of $(i)$ of Theorem \ref{fin}.}
Let the conditions of Theorem \ref{fin} be fulfilled.
   By Lemmas \ref{b-s}  we have
$$ N(z)=n(1,T(z))\,\,\mbox{as}\,\,z<0 $$ and by Lemma \ref{G-S}
for any $\gamma\in [0,1)$ the number $n(1-\gamma,T(0)) $ is finite.
Then for all $z<0$ and $\gamma \in (0,1)$ we have $$
N(z)=n(1,T(z))\leq n(1-\gamma,T(0))+n(\gamma,T(z)-T(0)). $$

This relation can be easily obtained by use of the Weyl inequality
$$
n(\lambda_1+\lambda_2,A_1+A_2)\leq
n(\lambda_1,A_1)+n(\lambda_2,A_2)
$$
for sum of compact operators $A_1$ and $A_2$ and for any positive
numbers $\lambda_1$ and $\lambda_2.$

Since $T(z)$ is continuous from the left up to $z=0$, we obtain $$
\lim_{z\to 0} N(z)= N(0)\leq n(1-\gamma,T(0))\,\, \mbox{for all}\,\,
\gamma \in (0,1). $$

Thus
 $$N(0)\leq n(1-\gamma,T(0))<\infty.$$ The latter
inequality proves the assertion (i) of Theorem \ref{fin}. \qed

\section{Asymptotics for the number of
eigenvalues of the  operator $H$.}

In this section we shall closely follow  A. Sobolev's method
\cite{Sob} to derive the asymptotics for the number of eigenvalues
of $H$.

We shall first establish the asymptotics of $n(1,T(z))$ as $z\to
-0.$ Then assertion $(ii)$ of Theorem \ref{fin} will be deduced by
a perturbation argument based on the following lemma.

 \begin{lemma}\label{comp.pert}
 Let $A (z)=A_0 (z)+A_1 (z),$ where $A_0(z)$ $(A_1(z))$ is
compact and continuous in $z<0$ $(z\leq 0).$  Assume that for some
function $f(\cdot),\,\, f(z)\to 0,\,\, z\to -0$ the limit $$
\lim_{z\rightarrow -0}f(z)n(\lambda,A_0 (z))=l(\lambda), $$ exists
and is continuous in $\lambda>0.$ Then the same limit exists for
$A(z)$ and $$ \lim_{z\rightarrow -0}f(z)n(\lambda,A
(z))=l(\lambda). $$
\end{lemma}
For the proof of Lemma \ref{comp.pert}, see Lemma 4.9 of
\cite{Sob}.

By Hypothesis \ref{hypoth u} we get
\begin{equation}\label{asymp1}
 u(p,q)=\frac{1}{2}\big(
l_1(Up,p)+2l(Up,q)+l_2(Uq,q))+|p|^4+|q|^4)\,\, \text{as}\,\,
p,q\rightarrow 0,
\end{equation}
and by the \eqref{malfa} and Corollary \ref{razl.lemma.natijasi.}
for any sufficiently small negative $z$ we get
\begin{equation}\label{asymp2}
\Delta_{\mu^0_\alpha}(p,z) = \frac{4\pi^2 \mu^0_\alpha
\varphi^2_\alpha(0)}
 {l_\beta^{{3}/{2}} \det(U)^{\frac{1}{2}}}
 \left [ n_\alpha (Up,p) -2z \right ]^{\frac{1}{2}}+
 O(|p|^2+|z|)\,\, \text{as}\,\, p,z\rightarrow 0, \end{equation}
where
\begin{equation*}
n_\alpha={(l_1l_2-l^2)}/{l_\beta}.
\end{equation*}

 Let $T(\delta;|z|)$ be an
operator on $L_2^{(2)}({\T}^3)$ with the entries
\begin{align*}
&T_{11}(\delta;|z|)=T_{22}(\delta;|z|)=0,\\ &(T
_{\alpha\beta}(\delta;|z|)w)(p)\\ &= \mathrm{d}_0
\int\limits_{\T^3} \frac{ \hat \chi_\delta (p) \hat \chi_\delta
(q) (n_\alpha(Up,p)+ 2|z|)^{-1/4} (n_\beta  (Uq,q)+ 2|z|)^ {-1/4}
} {l_\alpha(Up,p)+ 2l(Up,q)+l_\beta (Uq,q)
 +2|z|} w(q)d q,\\
 &w\in L_2(\T^3),\,\, \alpha\neq \beta,\,\, \alpha,\beta=1,2,
\end{align*}
where $\hat \chi_\delta(\cdot)$ is the  characteristic function of
$\hat U_\delta(0)=\{ p\in \T^3:\,\, |U^{\frac{1}{2}}p|<\delta \}$
and
 $$ \mathrm{d}_0= \frac{{\det U}^{\frac{1}{2}}} {2 \pi^2}l_1^{\frac{3}{4}}
l_2^{\frac{3}{4}}.
 $$

The main technical point to apply Lemma \ref{comp.pert} is the
following
\begin{lemma}\label{H-SH} Assume Hypotheses \ref{hypoth u},
\ref{hyp.varphi} and \ref{hypothD} are fulfilled and let $\mu_\alpha
=\mu_\alpha^0$ for all $\alpha=1,2.$  The operator $ T(z)-T(\delta;
|z|)$ belongs to the Hilbert-Schmidt class and is continuous in
 $z\leq 0.$
\end{lemma}
\begin{proof}
Applying the  asymptotics \eqref{asymp1}, \eqref{asymp2} and Lemmas
\ref{D.ineq} and \ref{U.ineq} one can estimate the kernel of the
operator $ T_{\alpha\beta} (z)-T_{\alpha \beta}(\delta; |z|)$ by
\begin{equation*}
 C [ (p^2+q^2)^{-1} +
|p|^{-\frac{1}{2}}(p^2+q^2)^{-1} +
|q|^{-\frac{1}{2}}(p^2+q^2)^{-1}+1 ]
\end{equation*}
 and hence
the operator $ T_{\alpha\beta} (z)-T_{\alpha \beta}(\delta; |z|)$
 belongs to the Hilbert-Schmidt class for all
$z \leq 0.$ In combination with the continuity of the kernel of
the operator in  $z<0$ this  gives   the continuity of $
T(z)-T(\delta;|z|)$ in  $z\leq 0.$
\end{proof}

Let us now recall some results from \cite{Sob}, which are
important in our work.\\
 Let ${\bf S}_{{\bf r}}:L_2((0,{\bf r})\times {\bf \sigma}^{(2)})\to
 L_2((0,{\bf r})\times {\bf \sigma}^{(2)})$ be the integral operator with the
kernel
\begin{equation}\label{sobolev}
 S_{\alpha\alpha}(y;t)=0,\quad
 S_{\alpha\beta}(y;t)=(2\pi)^{-2}\frac{u_{\alpha\beta}}
{\cosh(y+r_{\alpha \beta})+s_{\alpha\beta}t}
\end{equation}
and
\begin{align*}
u_{\alpha\beta}=u_{\beta\alpha}=\big( \frac{l_1l_2}{l_1l_2-l^2}
\big)^{\frac{1}{2}},\,
 r_{\alpha\beta}=\frac{1}{2} \log
\frac{l_\alpha}{l_\beta} ,\,
s_{\alpha\beta}=s_{\beta\alpha}=\frac{l} {\sqrt{l_1l_2}},\,\,
\alpha\neq \beta,\,\, \alpha,\beta=1,2,
\end{align*}
${\bf r}=1/2 | \log |z||,\, y=x-x',\,t=<\xi, \eta>,\,\xi, \eta \in
\S^2,\,{\bf \sigma}=L_2(\S^2),\,\S^2$ being unit sphere in $\R^3.$

 Let $\hat {\bf S}(y),\,y\in \R$ be the integral operator on ${\bf \sigma}^{(2)}$
 whose kernel depends on the scalar product $t=<\xi,\eta>$ of the arguments
$\xi,\eta\in\S^2$ and has the form
\begin{equation*}
\hat S_{\alpha\alpha}(y)=0,\quad \hat
S_{\alpha\beta}(y)=(2\pi)^{-2}\frac{u_{\alpha\beta}
e^{ir_{\alpha\beta}y} \sinh[y(arccos s_{\alpha\beta}t)]}
{\sqrt{1-s_{\alpha\beta}^2t}\sinh(\pi x)}.
\end{equation*}

For $\lambda>0,$ define
$$
{U}(\lambda)= (4\pi)^{-1} \int\limits_{-\infty}^{+\infty}
n(\lambda,\hat{\bf S}(y))dy
$$
and denote $\cU_0=U(1).$

The following lemma can be proved in the same way as Theorem 4.5 in
\cite{Sob}.
\begin{lemma}\label{lim} The following equality
$$ \lim\limits_{{\bf r}\to \infty} \frac{1}{2}{\bf
r}^{-1}n(\lambda,{\bf S}_{\bf r})={U}(\lambda) $$ holds.
\end{lemma}

 The following theorem is basic for the proof of the asymptotics
\eqref{asym.K}.
\begin{theorem}\label{main} The equality
$$ \lim\limits_{|z|\to 0} \frac{n(1,T(z))} {|log|z||}
=\lim\limits_{{\bf r}\to \infty} \frac{1}{2}{\bf r}^{-1}n(1,{\bf
S}_{\bf r}) $$ holds.
\end{theorem}

\begin{remark} Since $\cU(\cdot)$ is continuous in $\lambda,$ according
to Lemma \ref{comp.pert}  any perturbations of the operator
$A_0(z)$ defined in Lemma \ref{comp.pert}, which are compact and
continuous up to $z=0$ do not contribute to the asymptotics
\eqref{asym.K}. During the proof of Theorem \ref{main} we use this
fact without further comments.
\end{remark}

{\bf Proof of Theorem \ref{main}.} As in Lemma \ref{H-SH}, it can
be shown that $ T(z)-T(\delta; |z|),\,z\leq 0,$ defines a compact
operator continuous in $z\le 0$  and it does not contribute to the
asymptotics \eqref{asym.K}.

 The space of vector-functions
$w=(w_1,w_2)$ with coordinates having support in $\hat
U_\delta(0)$
 is an invariant subspace
for the operator $T(\delta;|z|).$

Let  $\hat T_0(\delta;|z|)$ be the restriction of the operator $T
(\delta,|z|)$ to the subspace $L^{(2)}_2(\hat U_{\delta}(0)).$ One
verifies  that the operator $\hat T_0(\delta;|z|)$ is unitarily
equivalent to the following operator $T_0(\delta;|z|)$ in
$L^{(2)}_2(\hat U_{\delta}(0))$ with the entries
\begin{align*}
&T^{(0)}_{11}(\delta;|z|)=T^{(0)}_{22}(\delta;|z|)=0,\\ &(T^{(0)}
_{\alpha\beta}(\delta;|z|)w)(p)= \mathrm{d}_1 \int_{U_\delta(0)}
\frac{ (n_\alpha p^2+ 2|z|)^{-1/4} (n_\beta  q^2+ 2|z|)^ {-1/4} }
{l_\alpha p^2+ 2l(p,q)+l_\beta q^2 +2|z|} w(q)d q,\\ &w\in
L_2(U_\delta(0)),\,\,\,\alpha\neq \beta,\,\, \alpha,\beta=1,2,
\end{align*}
where
 $$ \mathrm{d}_1= ({2 \pi^2})^{-1}{l_1^{{3}/{4}}
l_2^{{3}/{4}}} .
 $$

Here the equivalence is performed by the unitary dilation $${\bf
Y}=diag\{Y_1,Y_2,\}:L^{(2)}_2( U_{\delta}(0))
 \to L^{(2)}_2(\hat U_{\delta}(0)),\quad
 (Y_\alpha f)(p)=f(U^{-\frac{1}{2}}p),\alpha=1,2.
 $$

The operator $T_0(\delta;|z|)$ is unitary equivalent
 to the operator $T_1(\delta;|z|)$ with entries
\begin{align*}
&T^{(1)}_{11}({\delta};|z|)=T^{(1)}_{22}({\delta};|z|)=0,\\
&(T^{(1)}_{\alpha\beta}({\delta};|z|)w)(p)=
 \mathrm{d}_1 \int_{U_r(0)}
\frac{  (n_\alpha p^2+ 2|z|)^{-1/4} (n_\beta  q^2+ 2|z|)^ {-1/4} }
{l_\alpha p^2+ 2l(p,q)+l_\beta q^2 +2} w(q)d q,\\ &w\in
L_2(U_r(0)),\,\,\,\alpha\neq \beta,\,\, \alpha,\beta=1,2,
\end{align*}
acting in $L_2^{(2)}(U_r(0)),\,U_r(0)=\{ p\in \R^3:|p|<r\},\,\,
r=|z|^{-\frac{1}{2}}.$

The equivalence is performed by the unitary dilation
 $${\bf B}_r=diag\{B_r,B_r\}:L_2^{(2)}
(U_\delta(0)) \to L_2^{(2)}(U_r(0)),\quad
 (B_r f)(p)=(\frac{r}{\delta})^{-3/2}f(\frac{\delta}
{r}p).$$ Further, we may replace $$(n_\alpha p^2+2)^{-1/4},\,
(n_\beta q^2+2)^{-1/4} \quad \mbox{ and}\quad l_\alpha p^2+
2l(p,q)+l_\beta q^2
 +2$$
 by
$$(n_\alpha p^2)^{-1/4}(1-\chi_1(p)),\,\, (n_\beta
q^2)^{-1/4}(1-\chi_1(q))
 \quad \mbox{ and}\quad
l_\alpha p^2+ 2l(p,q)+l_\beta q^2 ,$$
  respectively, since the error will be
a Hilbert-Schmidt operator  continuous up to  $z=0$. Then we get
the operator $T_2(r)$ in $L_2^{(2)}(U_r(0) \setminus U_1(0))$ with
entries
\begin{align*}
&T^{(2)}_{11}(r)=T^{(2)}_{22}(r)=0,\\
&(T^{(2)}_{\alpha\beta}(r)w)(p)= (n_1 n_2)^{-\frac{1}{4}}
\mathrm{d}_1\int_{U_r(0)\setminus U_1(0)} \frac{|p|^{-1/2}|
q|^{-1/2}} {l_\alpha  p^2+ 2l(p,q)+l_\beta  q^2} w(q)dq,\\ &w\in
L_2(U_r(0) \setminus U_1(0)),\,\,\,\alpha\neq \beta,\,\,
\alpha,\beta=1,2.
\end{align*}

The operator $T_2(r)$ is unitarily equivalent to the integral
operator ${\bf S}_{{\bf r}}$ with entries \eqref{sobolev}. The
equivalence is performed by the unitary operator $${\bf M}=diag\{M,M
\}:L_2^{(2)}(U_r(0) \setminus U_1(0)) \longrightarrow L_2((0,{\bf
r})\times {\bf \sigma}^{(2)}),$$
 where
$(M\,f)(x,w)=e^{3x/2}f(e^{ x}w), x\in (0,{\bf r}),\, w \in
{\S}^2.$ \qed

{\bf Proof of $(ii)$ of Theorem \ref{fin}} Similarly to \cite{Sob}
we can show that
\begin{equation}\label{Slambda}
\cU_0={U}(1) \ge \frac{1}{4\pi} \int\limits_{-\infty}^{+\infty}
n(1,\hat{\bf S}^{(0)}(y))dy\ge\frac{1}{4\pi}
mes\{y:2u_{12}e^{-\frac{\pi |y|}{2}}>1\},
\end{equation}
where $\hat{\bf S}^{(0)}(y),\, y\in \R$  is the $2\times2-$ matrices
with the entries
\begin{equation*}
\hat {S}^{(0)}_{\alpha\beta}(y)=\frac{u_{\alpha\beta}
e^{ir_{\alpha\beta}y}\sinh(y arcsin s_{\alpha\beta})}
{s_{\alpha\beta}y\cosh \frac{\pi y}{2}}
\end{equation*}
in the subspace of the harmonics of degree zero (see \cite{Sob}).

The positivity of $\cU_0$ follows from the fact that $u_{12}>1$
and $\hat {S}^{(0)}_{\alpha\beta}(0)>1$ and the continuity of
$\hat {S}^{(0)}_{\alpha\beta}(y).$

 Taking into account the inequality
\eqref{Slambda} and Lemmas \ref{b-s}, \ref{main}, \ref{lim}, we
complete the proof of $(ii)$ of Theorem \ref{fin}. \qed
\appendix
\section{Some properties of the function $u^{(\alpha)}_p(q)$}

\begin{lemma}\label{minimum}
Let Hypothesis \ref{hypoth u} be fulfilled. Then there exists a $ {
\delta } $-neighborhood $U_ {\delta } (0)\subset \T^3$ of the point
$p=0$ and an analytic function $q_ { \alpha } (p)$ defined on $U_
{\delta } (0) $
  such
  that:\\
(i)   for any $p { \in } U_ { \delta }(0)$  the point
  $q_ { \alpha } (p) $ is
a unique
  non-degenerate minimum  of the function
$u_p^{(\alpha)}(q)$ and
\begin{equation*}\label{p.alpha.k}
q_\alpha(p)=-\frac{l}{l_\beta}p+O(|p|^3)\,\,as\,\,p \to 0.
\end{equation*}\\
(ii) The function $m_{\alpha}(p)= u_{p}(q_\alpha(p))$ is
analytic in $U_ {\delta } (  0 ) $ and satisfies 
 \begin{equation}\label{min.raz}
m_{\alpha}(p)=\frac{l_1 l_2-l^2}{2l_\beta}(Up,p)+O(|p|^4) \quad
\mbox{as}\quad p \to 0,\quad \alpha\neq \beta,\, \alpha,\beta=1,2.
\end{equation}
\end{lemma}
\begin{proof}
$(i)$ By Hypothesis \ref{hypoth u} we obtain $$
u_0^{(\alpha)}(q)>u_0^{(\alpha)}(0),\,q \neq 0 $$ and $$ \left(
\frac{\partial^2 u_0^{(\alpha)}(0)}{\partial q^{(i)}
\partial q^{(j)}} \right)_{i,j=1}^3= l_\alpha U.
$$

Since $U$ is a positive definite matrix  the function $
u_0^{(\alpha)}(q), { \alpha } =1,2,$
 has a unique non-degenerate  minimum at  $q=0,$
 the gradient
  $ { \bigtriangledown } u_0^{(\alpha)}(q) $
   is equal to  zero at the point $q=0.$

Now we apply the implicit function  theorem
      to the equation
     $$ {\bigtriangledown } u_p^{(\alpha)}(q)=0
,\,\,p,q\in {\T}^3.$$ Then there exists a $ { \delta }
$-neighborhood  $U_ { \delta } (0 ) $ of the point $p=0$ and a
vector function $q_\alpha (p)$  defined and analytic  in $U_ {
\delta }(0) $ and for all $p { \in } U_ { \delta } ( 0 ) $ the
identity $ { \bigtriangledown } u _p^{(\alpha)}  ( q_\alpha(p))
 \equiv 0 $
holds.

Denote by $B(p) $ the matrix of the second order partial
derivatives of the function $ u_p^{(\alpha)} ( q) $
 at the point
$ q_{\alpha}(p)$. The matrix $B(0)=l_\alpha U$ is positive and  $B
(p)$ is continuous in $U_ { \delta } ( 0 )$ and hence  for any $p
\in U_{\delta} ( 0 )$
 the matrix $B (p)$ is positive definite. Thus
$ q_{\alpha}(p),\,p \in U_ { \delta } ( 0 )$ is the unique
non-degenerate minimum point
 of $u_p^{(\alpha)} (q).$

The    non-degenerate  minimum point $q_\alpha(p)$ is an odd
function in $p \in U_\delta(0).$

Indeed, since $u(p,q)$ is even we get $u_p^{(\alpha)} (
-q)=u_{-p}^{(\alpha)} ( q)$, and   we obtain $$
u^{(\alpha)}_{-p}(-q_\alpha(p))= m_{\alpha}(p)=m_{\alpha}(-p)=
u^{(\alpha)}_{-p}(q_\alpha(-p)). $$

Since for all $p \in U_\delta(0)$ the point $q_\alpha(p)$ is the
unique non-degenerate  minimum of $u^{(\alpha)}_{p}(q)$ we have $$
q_\alpha(-p)=-q_\alpha(p). $$

By Hypothesis \ref{hypoth u} and the Taylor expansion
 we get
 \begin{equation}\label{taylor2}
u_p^{(\alpha)}(q -\frac{l}{l_\beta}p )=
 \frac{l_\beta}{2}(Uq,q)+
\frac{l_1l_2-l^2}{2l_\beta}(Up,p)+
 O(|q|^4+|p|^4)\,\,\mbox{as}\,\,q,p
\to 0.
\end{equation}

Since $u_p^{(\alpha)}(q_\alpha(p))=\min\limits_{q\in {\T}^3}
 u_p^{(\alpha)}
 (q ) \leq
u_p^{(\alpha)}(-\frac{l}{l_\beta}p)$ and $q_\alpha(p)$ is odd
 we have
$$
\frac{l_\beta}{2}\big(U^{\frac{1}{2}}(q_{\alpha}(p)+\frac{l}{l_\beta}p
) \big )^2+ \frac{l_1l_2-l^2}{2l_\beta}(Up,p)+
 O(|p|^4)\leq
 \frac{l_1l_2-l^2}{2l_\beta}(Up,p)+
 O(|p|^4)
 \,\,\mbox{as}\,\,p
\to 0
 $$
 that is,
$$l_2\big(U^{\frac{1}{2}}(q_{\alpha}(p)+\frac{l}{l_\beta}p ) \big
)^2 \leq
 O(|p|^4)
 \,\,\mbox{as}\,\,p \in U_\delta(0). $$

This inequality is not valid if $q_\alpha(p)$ has the asymptotics
$q_\alpha(p) + \frac{l}{l_\beta}p=O(|p|)\,\,as\,\,p \to 0.$ Since
$q_\alpha(p)$ is an odd analytic function,
 we have
$q_\alpha(p)+\frac{l}{l_\beta}p=O(|p|^3)\,\,as\,\,p \to 0.$

$(ii)$ Since the functions $u(p,q),\,p,q \in \T^3$ and
$q_\alpha(p),\, p\in U_{\delta}(0)$
 are analytic, we have that the function
$m_{\alpha}(p)=u^{(\alpha)}_p(q_\alpha(p))$
 is also
analytic on $p \in U_{\delta}(0) $.

By    $q_\alpha(p)=-\frac{l}{l_\beta}p+O(|p|^3),\,p \to 0$ and
\eqref{taylor2} we get the asymptotics \eqref{min.raz}.
\end{proof}

Denote by $U_\alpha(p,q)$ the function defined on
$U_{\delta}(0)\times \T^3$  as
 \begin{equation*}
 U_\alpha(p,q)=u^{(\alpha)}_p
(q+q_\alpha(p))-m_{\alpha}(p),
\end{equation*}
where the function $q_\alpha(p),\,p \in U_{\delta}(0)$ is defined
in
 Lemma \ref{minimum}.

\begin{lemma}\label{ras.u(p,q)} Let Hypothesis \ref{hypoth u} be fulfilled.
For any $p,q \in U_{\delta}(0)$ the function $U_\alpha(p,q)$  is
even and represented as \begin{equation*} U_\alpha(p,q)=
 \frac{l_\beta}{2}
(Uq,q) +h(p,q) ,
\end{equation*}
 where $h(p,q)$ satisfies $ h(p,q)=h(-p,-q)$ and
\begin{equation}\label{ex1}
h(p,q)=O(|p|^2|q|^2)+O(|p||q|^3)+O(|q|^4) \,\,as\,\,|p|,|q| \to 0.
\end{equation}
\end{lemma}
\begin{proof} Since the functions $m_\alpha(p)$, $q_\alpha(p)$ are analytic in
$p \in U_{\delta}(0) $ and $u(p,q)$ is real
 analytic on $(\T^3)^2$ we have that
the function $ U_\alpha(p,q)$  is  analytic in $(p,q)\in
U_{\delta}(0)\times \T^3$.

Using the representation  \eqref{taylor2} and Lemma \ref{minimum}
we have \begin{equation}\label{ex2} U_\alpha(p,q)-
 \frac{l_\beta}{2}(Uq,q) =h(p,q),\quad  h(p,q)=O(|p|^4+|q|^4) \,\,
\mbox{as}\,\,p,q\to 0.
 \end{equation}

From the equality $u_{-p}^{(\alpha)}(-q)=u_{p}^{(\alpha)}(q)$
 and  oddness of $q_\alpha(p)$ we obtain evenness
 of $U_\alpha(p,q)$, that is, $U_\alpha(p,q)=U_\alpha(-p,-q)$
 for all  $p \in U_{\delta}(0),\,q \in \T^3$.
 Then we have $ h(p,q)= h(-p,-q).$

By Lemma \ref{minimum}  for all $p\in U_{\delta}(0)$ the point
$q_\alpha(p)$ is the minimum point of $u_{p}^{(\alpha)}(q)$ we get
from \eqref{ex2} that  $ h(p,0)=0$ and
 $\triangledown
 h(p,0)=
  ( \frac  { \partial  h(p,0) }
   {\partial  q^{(1)}},
\frac { \partial  h(p,0) }
 { \partial
q^{(2)}},
 \frac { \partial
 h(p,0)} { \partial q^{(3)} }) =0. $

Since $ h(p,q)$ is analytic the Taylor expansion around the point
$(0,0)$ gives \eqref{ex1}. \end{proof}
\begin{lemma}\label{U.ineq} Let Hypotheses \ref{hypoth u} be
fulfilled. Then there exist numbers $C_1, C_2,C_3>0$ and
$\delta>0$ such that the following inequalities
\begin{align*}
&(i)\quad C_1 (|p|^2+|q|^2)\leq u(p,q) \leq C_2 (|p|^2+|q|^2)\quad
\mbox{for all} \quad p, q\in U_\delta(0),\\
 &(ii)\quad u(p,q) \geq C_3\quad \mbox{for all} \quad(p, q)\notin
U_\delta(0)\times U_\delta(0)
\end{align*}
hold.
\end{lemma}
\begin{proof}
 By Hypothesis \ref{hypoth u} the point $(0,0)\in (\T^3)^2$ is
a unique non-degenerated minimum point of $u(p,q).$ Then by
\eqref{taylor2} there exist positive  numbers $C_1,C_2,C_3$ and a
$\delta-$neighborhood of $p=0\in \T^3$ so that $(i)$ and $(ii)$
hold true.
\end{proof}

\section{}

\begin{lemma}\label{examp}
Let
\begin{equation*}
 u(p,q)=\sum_{i=1}^3\big (
3-cos p^{(i)}-cos q^{(i)}-cos( p^{(i)}- q^{(i)})
 \big ),\quad
\end{equation*}
and either
\begin{equation}\label{U.cos}
\varphi_\alpha(p)= a_\alpha^{(0)}+\sum_{i=1}^3a_\alpha^{(i)} \cos
p^{(i)},\quad a_\alpha^{(j)}\in \R^1,\,j=0,1,2,3
\end{equation}
or
\begin{equation}\label{U.sin}
\varphi_\alpha(p)=a_\alpha\sum_{i=1}^3 \sin p^{(i)},\quad
a_\alpha\in \R^1.
\end{equation}
Then Hypotheses \ref{hypoth u},\ref{hyp.varphi} and \ref{hypothD}
are  fulfilled.
\end{lemma}
\begin{proof}
It is easy to see that Hypotheses \ref{hypoth u} and
\ref{hyp.varphi} are fulfilled.

We prove that  Hypothesis \ref{hypothD} is fulfilled. Since
$u(p,q)$ and $|\varphi_\alpha(p)|$ are even the function
$\Lambda_{\alpha}(p)$  is also even.

Then we get
\begin{equation}\label{L.Even}
\Lambda_{\alpha}(p)-\Lambda_{\alpha}(0)= \int\limits_{{\T}^3}
\frac{2u_0^{(\alpha)}(t)-(u_p^{(\alpha)}(t)+u_{-p}^{(\alpha)}(t))}
{4u_p^{(\alpha)}(t)u_{-p}^{(\alpha)}(t)u_0^{(\alpha)}(t)}[u_p^{(\alpha)}(t)+
u_{-p}^{(\alpha)}(t)]\varphi_\alpha^2(t)dt-
\end{equation}
$$
-\frac{1}{4}\int\limits_{{\T}^3} \frac{ [u_p^{(\alpha)}(t)-
u_{-p}^{(\alpha)}(t)]^2} {u_p^{(\alpha)}(t)
u_{-p}^{(\alpha)}(t)u_0^{(\alpha)}(t)}\varphi_\alpha^2(t)dt.
$$

From the equality
\begin{equation*}
u_0^{(\alpha)}(t)-\frac{u_p^{(\alpha)}(t)+u_p^{(\alpha)}(-t)}{2}=
\sum_{j=1}^{3}(\cos p^{(i)}-1)(1+\cos t^{(i)})
\end{equation*}
and \eqref{L.Even}  we get for all $p\in \T^3,\,p\neq 0$ the
inequality
\begin{equation*}
\Lambda_{\alpha}(p)-\Lambda_{\alpha}(0)= \int\limits_{{\T}^3}
\frac{ \sum_{j=1}^{3}(\cos p^{(i)}-1)(1+\cos t^{(i)})
\varphi_\alpha^2(t)dt}{u_p^{(\alpha)}(t)u_p^{(\alpha)}(-t)u_0^{(\alpha)}(t)}<0,
\end{equation*}
that is, the assertion  $(i)$ of Hypothesis \ref{hypothD} holds.

Since for any $p,q\in \T^3,p\neq 0$ the inequality
$u_p^{(\alpha)}(q)>0$ holds for any nonzero $p\in \T^3$ and
$\varphi_\alpha(0)=0$ the integrals
\begin{equation*}\label{L.Def}
\lambda_{ij}^{(1)}(p)=\int\limits_{{\T}^3} \frac{ \frac{\partial^2
} {\partial p^{(i)}
\partial p^{(j)}
}u_p^{(\alpha)}(t) \varphi_\alpha^2(t)dt}{(u_p^{(\alpha)}(t))^2}
\end{equation*}
and
\begin{equation*}\label{L.Def2}
\lambda_{ij}^{(2)}(p)=2\int\limits_{{\T}^3} \frac{ \frac{\partial
} {\partial p^{(i)} }u_p^{(\alpha)}(t) \frac{\partial } {\partial
p^{(j)} }u_p^{(\alpha)}(t)
\varphi_\alpha^2(t)dt}{(u_p^{(\alpha)}(t))^3},\quad i,j=1,2,3,
\end{equation*}
are finite and hence are bounded continuous functions on $\T^3.$

Then the function $\Lambda_{\alpha}(p)$ is a twice differentiable
function on $\T^3$ and
\begin{equation*}\label{L.Def3}
\frac{\partial^2 \Lambda_{\alpha}(p)} {\partial p^{(i)}
\partial p^{(j)}}=-\lambda_{ij}^{(1)}(p)+\lambda_{ij}^{(2)}(p),
\quad i,j=1,2,3.
\end{equation*}

Since
\begin{align*}\label{deff}
&  \frac{\partial } {\partial p^{(i)} }u_0^{(\alpha)}(t) =\sin
t^{(i)},\,\,  \frac{\partial^2 } {\partial p^{(i)}\partial p^{(i)}
}u_0^{(\alpha)}(t) =1+\cos t^{(i)},\\
& \frac{\partial^2 } {\partial p^{(i)}\partial p^{(j)}
}u_0^{(\alpha)}(t)=0,\quad i\neq j,\,\,i,j=1,2,3,\nonumber
\end{align*}
we get
\begin{equation*}\label{L.Def4}
\frac{\partial^2 \Lambda_{\alpha}(0)} {\partial p^{(i)}
\partial p^{(i)}}=-
2\int\limits_{{\T}^3} \frac{ \sum_{s=1,s\neq i}^3 (1-\cos
t^{(s)})(1+\cos t^{(i)})
\varphi_\alpha^2(t)dt}{(u_0^{(\alpha)}(t))^3} ,
\end{equation*}
\begin{equation*}\label{L.Def5}
\frac{\partial^2 \Lambda_{\alpha}(0)} {\partial p^{(i)}
\partial p^{(j)}}=2
\int\limits_{{\T}^3} \frac{ \sin t^{(i)}\sin t^{(j)}
\varphi_\alpha^2(t)dt}{(u_0^{(\alpha)}(t))^3},\quad i\neq j,\,
i,j=1,2,3.
\end{equation*}

If  $\varphi_\alpha(p) $ is defined by \eqref{U.cos}, then from
the last two equalities we get
\begin{align}\label{par.der0}
&\frac{\partial^2 \Lambda_{\alpha}(0)} {\partial p^{(i)}
\partial p^{(i)}}<0,
\frac{\partial^2 \Lambda_{\alpha}(0)} {\partial p^{(i)}
\partial p^{(j)}}=0,\,\,i\neq j,\,\,i,j=1,2,3.
\end{align}

If  $\varphi_\alpha(p) $ is defined by \eqref{U.sin} then
\begin{align}\label{par.der}
&\frac{\partial^2 \Lambda_{\alpha}(0)} {\partial p^{(1)}
\partial p^{(1)}}=\frac{\partial^2 \Lambda_{\alpha}(0)} {\partial p^{(2)}
\partial p^{(2)}}=\frac{\partial^2 \Lambda_{\alpha}(0)} {\partial p^{(3)}
\partial p^{(3)}}<0,\\
&\frac{\partial^2 \Lambda_{\alpha}(0)} {\partial p^{(1)}
\partial p^{(2)}} =\frac{\partial^2 \Lambda_{\alpha}(0)} {\partial p^{(1)}
\partial p^{(3)}}=
\frac{\partial^2 \Lambda_{\alpha}(0)} {\partial p^{(2)}
\partial p^{(3)}}>0,\nonumber\\
&\frac{\partial^2 \Lambda_{\alpha}(0)} {\partial p^{(i)}
\partial p^{(i)}}+2\frac{\partial^2 \Lambda_{\alpha}(0)} {\partial p^{(i)}
\partial p^{(j)}}<0,\quad
i\neq j,\,\,i,j=1,2,3.\nonumber
\end{align}

Using \eqref{par.der0} (resp.
 \eqref{par.der}) we get that the
matrix of the second order partial derivatives of the function $
\Lambda_{\alpha}(p) $
 at the point
$p=0$ is negative definite.

Thus the function $\Lambda_{\alpha}(p)$ has a non-degenerate maximum
at the point $p=0.$
\end{proof}

{\bf Acknowledgement} The authors would like to thank Prof. R.A.
Minlos and Dr. Z.I.Muminov  for several helpful discussions about
 the results of the paper. This work was
supported by the DFG 436 USB 113/4 Project and the Fundamental
Science Foundation of Uzbekistan. S.N.Lakaev  gratefully
acknowledge the hospitality of the Institute of Applied
Mathematics and of the IZKS of the University of Bonn.

\end{document}